\documentclass[
 reprint,
 amsmath,
 pre,
 amssymb,
 nofootinbib,
 aps,
 longbibliography,
]{revtex4-1}
\usepackage[pdftex]{graphicx}
\usepackage{wasysym}
\usepackage[bottom]{footmisc}
\usepackage{amsfonts}
\usepackage{ifsym}
\usepackage{pifont}
 \usepackage{footmisc}
\usepackage[export]{adjustbox}
\usepackage{dsfont}
\usepackage{lipsum}
\usepackage{stmaryrd}
\usepackage{MorrisIn,lettrine,Romantik}

\usepackage{subcaption}

\usepackage{tikz}
\usepackage{tikzit}

\tikzstyle{cyl}=[shape=cylinder, draw=none, minimum height=20mm, minimum width=10mm, cylinder uses custom fill, cylinder body fill={rgb,255: red,135; green,192; blue,241}, cylinder end fill={magenta!40}, tikzit draw=blue, inner sep=0]
\tikzstyle{Wbar}=[draw=black, shape=ellipse, minimum height=15mm, minimum width={15mm*1/6}, fill={rgb,255: red,146; green,190; blue,241}, semithick]
\tikzstyle{rectangle}=[fill={rgb,255: red,106; green,167; blue,241}, draw=none, shape=rectangle, minimum height=15mm, minimum width=20mm]
\tikzstyle{S1}=[fill=none, draw=black, shape=ellipse, minimum height=15mm, minimum width={15mm*1/6}, dashed, dash pattern=on 2.5pt off 2pt]
\tikzstyle{W}=[fill={rgb,255: red,106; green,167; blue,241}, draw=black, shape=ellipse, minimum height=15mm, minimum width={15mm*1/6}, semithick, dash pattern=on 45 off 45, dash phase=-22.5]
\tikzstyle{origin}=[fill={rgb,255: red,237; green,0; blue,0}, draw=none, shape=circle, radius=0.1mm, inner sep=0pt]
\tikzstyle{Knode}=[fill=none, draw=none, rotate=-90, shape=circle]
\tikzstyle{pnode}=[fill=none, draw=none, shape=circle, rotate=-45]

\tikzstyle{arrowhead}=[draw=black, ->, semithick]
\tikzstyle{R1}=[-, dash pattern=on 2.5pt off 2pt]
\tikzstyle{arrow}=[thin, <->]
\tikzstyle{bdy}=[-, fill={rgb,255: red,146; green,190; blue,241}, draw=none]
\tikzstyle{axis}=[->, line width=0.2mm]
\tikzstyle{cut}=[-, draw={rgb,255: red,237; green,0; blue,0}, line width=0.5mm]
\tikzstyle{B}=[-, line width=0.5mm]
\tikzstyle{int}=[-, fill={rgb,255: red,213; green,213; blue,213}, draw=black]
\tikzstyle{legarrow}=[->]

\let\oldaddcontentsline\addcontentsline

\newcommand{\starttocentries}{\let\addcontentsline\oldaddcontentsline}
\makeatletter
\newlength{\apb@width}
\newcommand{\autoparbox}[2][c]{\settowidth{\apb@width}{#2}\parbox[#1]{\apb@width}{#2}}

\makeatother

\newcommand{\namedref}[2]{\hyperref[#2]{#1~\ref*{#2}}}

\newcommand{\br}[1]{\langle#1\rangle}

\newcommand{\eps}{\varepsilon}

\newcommand{\Csphere}{{}^\bullet\kern-1.2pt C}
\newcommand{\Ctorus}{{}^\circ\kern-1.2pt C}

\newcommand{\COMMENT}[1]{}

\newcommand{\neqa}{\nonumber\end{eqnarray}}

\newcommand{\<}{{\langle}}
\renewcommand{\>}{{\rangle}}

\newcommand{\re}{\relax{\rm I\kern-.18em R}}

\def\su2{{SU(2)}}

\def\eps{{\epsilon}}
\def\o{{\omega}}

\def\[{\left[}
\def\]{\right]}

\def\({\left(}
\def\){\right)}
\def\[{\left[}
\def\]{\right]}

\def\<{\langle}
\def\>{\rangle}

\def\i2{\frac{i}{2}}

\def\2F1{\,_2{\rm F}_1}

\usepackage[usenames,dvipsnames]{xcolor}
\usepackage{color}
\usepackage{graphicx}
\usepackage{dcolumn}
\usepackage{bm}

\usepackage{cancel}
\usepackage{ctable}
\usepackage{booktabs}
\newcolumntype{L}[1]{>{\raggedright\let\newline\\\arraybackslash\hspace{0pt}}m{#1}}
\newcolumntype{C}[1]{>{\centering\let\newline\\\arraybackslash\hspace{0pt}}m{#1}}
\newcolumntype{R}[1]{>{\raggedleft\let\newline\\\arraybackslash\hspace{0pt}}m{#1}}

\newcommand{\beq}{\begin{equation}}
\newcommand{\eeq}{\end{equation}}
\newcommand{\beqq}{\begin{equation*}}
\newcommand{\eeqq}{\end{equation*}}
\newcommand\beqa{\begin{eqnarray}}
\newcommand\eeqa{\end{eqnarray}}
\newcommand\beqaa{\begin{eqnarray*}}
\newcommand\eeqaa{\end{eqnarray*}}
\newcommand\bea{\begin{array}}
\newcommand\eea{\end{array}}

\usepackage[hidelinks]{hyperref}

\begin{document}

\setlength{\skip\footins}{0.5cm}
\addtolength{\textheight}{0.155in}  
\addtolength{\topmargin}{-0.06in}

\title{
Imprints of asymptotic freedom on confining strings
}

\author{Jan Albert$^{a}$, Alexandre Homrich$^{b,c}$} 
\affiliation{$^{a}$Princeton Center for Theoretical Science, Princeton University, Princeton, NJ 08544, USA}
\affiliation{\vspace{0.1cm} $^{b}$ Kavli Institute for Theoretical Physics,
University of California, Santa Barbara, CA 93106, USA
\\$^{c}$ Walter Burke Institute for Theoretical Physics, Caltech, Pasadena, CA 91125, USA}
             
\begin{abstract}
We consider the Polyakov loop correlator in the confining phase of large $N$ Yang-Mills theory in three and four dimensions. It can be computed by summing over the exchange of closed flux tubes winding around the thermal cycle. At short separations, the leading divergence is controlled by perturbation theory. Combining these two facts allows us to determine the asymptotic spectral density of string states contributing to the correlator. This sharply relates the weakly-coupled UV of the gauge theory to the dynamics of highly energetic flux tubes.
Then, in a toy integrable setting, we explore how this
can bound the scattering data of the Goldstone modes on top of a long string. We derive a bound on the asymptotic behavior of the reflection amplitude  of Goldstones against the flux tube boundary sourced by the Polyakov line, and rule out an asymptotically linear phase shift for the S-matrix. 
Along the way, we discuss how causality can impose bounds on thermodynamic quantities, and show how the positivity of time delays follows from unitarity and analyticity of $2d$ massless elastic S-matrices. We include a review on reflection amplitudes, and their computation in the theory of long effective strings.

\end{abstract}

\pacs{Valid PACS appear here}

\hfill CALT-TH 2026-007

\maketitle

\section{Introduction}
  \lettrine{U}{nderstanding} confinement requires connecting the dynamics of almost-free gluons at high-energies to the behavior of emergent confining flux tubes sourced by far separated probe quarks. In this letter, we describe some of the simplest consequences of asymptotic freedom \cite{Gross:1973id,Politzer:1973fx} for the high-energy dynamics of confining strings in large $N$ pure Yang-Mills theory in three and four dimensions.

    It is widely believed that this connection will be realized by recasting the gauge theory as a theory of strings~\cite{tHooft:1973alw, Nambu:1974zg,Polyakov:1987hqn}. While such a reformulation has been achieved for a variety of gauge theories by means of the AdS/CFT correspondence \cite{Maldacena:1997re,Gubser:1998bc,Witten:1998qj}, it remains unavailable for pure Yang-Mills (see \cite{Gross:1993hu,Aharony:2023tam,Komatsu:2025sqo,Aharony:2025owv,Komatsu:2025dqv} for progress in 2D).
    
    Toward this goal, it has proven useful to consider first a long confining string.
    The low-energy structure of the worldsheet is then quite rigid.
 This limit is described by $D-2$ massless transverse string modes $X^i$ whose interactions are constrained by the non-linearly realized spacetime Poincaré symmetry \cite{Nambu:1974zg,Luscher:1980ac,Polchinski:1991ax,Dubovsky:2012sh,Aharony:2013ipa}.
 The Goldstone bosons are free to leading order, and their S-matrix is entirely fixed up to order $(\ell_s E)^6$ in terms of the string tension $\ell_s^{-2}$. This is clear from the EFT action, which organizes in terms of curvature invariants \cite{Isham:1971dv,Volkov:1973vd, Polyakov:1986cs, Kleinert:1986bk}, schematically,
 \begin{equation}\label{eq:SEST}
    S_{\text{EST}} = -\int_{\Sigma} d^2\sigma \, \sqrt{-\det \partial X\cdot \partial X} \left(\frac{1}{\ell_s^2} +   K^4+\cdots\right)\,. 
\end{equation}
 The next several orders of the S-matrix are determined in terms of only a handful of non-universal Wilson coefficients. These and other low-energy observables can be rigorously bounded by bootstrap techniques \cite{Paulos:2016but,EliasMiro:2019kyf,EliasMiro:2021nul,Gaikwad:2023hof,Guerrieri:2024ckc}. 

 Both the effective theory analysis and most S-matrix bootstrap bounds are agnostic to the UV completion and thus apply to most theories with long, stable, string-like excitations. 
These are very diverse and include, beyond the chromoelectric flux tubes, domain walls in the Ising model \cite{Caselle:2002ah, Baffigo:2023rin, Lima:2025sqa}
and vortex strings in the Abelian Higgs model \cite{Abrikosov:1956sx, Nielsen:1973cs}. Without extra input, it is therefore natural that these methods have almost nothing to say about the dynamics at energy scales larger than $\ell_s^{-1}$.

A glimpse at the Yang-Mills flux tube structure at higher energies is provided by lattice simulations \cite{Luscher:2002qv,Athenodorou:2011rx,Caselle:2024zoh, Sharifian:2025fyl}.
The primary observables are the energy levels of closed winding strings, sourced by Polyakov loops in $\mathbb{R}^{D-1} \times S^1_R$, as a function of the circle radius $R$. The resulting finite volume spectrum clearly shows the existence of a light-massive pseudoscalar resonance in 4D, while in 3D no signature of resonances is manifest \cite{Dubovsky:2013gi}. This light-resonance spectrum has been speculated to reflect the high-energy structure of the theory, with the quantum numbers of the Goldstones and the axion (or absence thereof in 3D) relating to the low-dimension operator insertions in the Polyakov line \cite{Dubovsky_2018, Gabai:2025hwf}.  

Turning these qualitative expectations into sharp analytic predictions is nontrivial. The key challenge is to find flux tube observables whose high-energy limit is genuinely controlled by asymptotic freedom and thus accessible in weakly coupled Yang–Mills. Neither the worldsheet S-matrix nor the finite-volume spectrum do the job:
scattering of high-energy Goldstones occurs on top of a long string, thus making it hard to factorize the UV data from the IR background \cite{Dubovsky_2018}, while trying to probe short distances by shrinking $R$ eventually triggers a transition out of the confining phase in which the string description applies \cite{Polyakov:1978vu}.

\begin{figure}[t]
\centering
\scalebox{1.5}{\begin{tikzpicture}
	\begin{pgfonlayer}{nodelayer}
		\node [style=rectangle] (0) at (0, 0) {};
		\node [style=Wbar] (2) at (2, 0) {};
		\node [style=none] (3) at (2, 1.5) {};
		\node [style=none] (4) at (2, -1.5) {};
		\node [style=none] (5) at (-2, -1.5) {};
		\node [style=none] (6) at (-2, 1.5) {};
		\node [style=W] (7) at (-2, 0) {};
		\node [style=S1] (8) at (3, 0) {};
		\node [style=S1] (9) at (-3.25, 0) {};
		\node [style=none] (10) at (3, 1.5) {};
		\node [style=none] (11) at (3, -1.5) {};
		\node [style=none] (12) at (-3.25, 1.5) {};
		\node [style=none] (13) at (-3.25, -1.5) {};
		\node [style=none] (14) at (2, -2.25) {$W_{\overline{\Box}}$};
		\node [style=none] (15) at (-2, -2.25) {$W_{\Box}$};
		\node [style=none] (16) at (-2, 1.8) {};
		\node [style=none] (17) at (2, 1.8) {};
		\node [style=none] (20) at (0, 2.25) {$\tau$};
		\node [style=none] (21) at (3.75, 0) {$R$};
	\end{pgfonlayer}
	\begin{pgfonlayer}{edgelayer}
		\draw [style=R1] (3.center) to (10.center);
		\draw [style=R1] (4.center) to (11.center);
		\draw [style=R1] (13.center) to (5.center);
		\draw [style=R1] (12.center) to (6.center);
		\draw [style=arrow, in=180, out=0] (16.center) to (17.center);
	\end{pgfonlayer}
\end{tikzpicture}}
\caption{The correlator of two Polyakov loops is equivalent to the worldsheet cylinder partition function.}
\label{fig:Cylinder}
\end{figure}
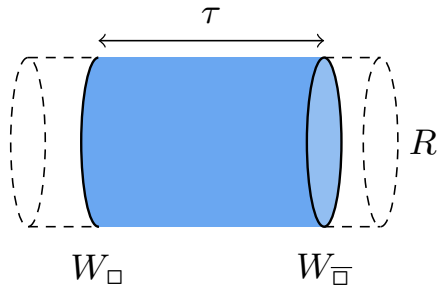

In this letter we point out that the two-point Polyakov loop correlator fits the bill.
This is, of course, the same object used in the lattice to extract finite-volume spectra, see figure \ref{fig:Cylinder} for an illustration.
At large $\tau, R$ it is described by the long string EFT, while its finite $R$, small $\tau$ asymptotics is controlled by perturbative Yang–Mills. Importantly, the correlator is a smooth function of $\tau$, with no phase transition, and thus interpolates between weakly coupled gluons and long confining strings. We discuss this in more detail in section~\ref{sec:IIB}.

At finite $N$, this interpolation is not captured by winding worldsheet states alone, as around $\tau \sim \ell_s^{-1}$, intermediate glueballs start to give important contributions. In the strict large $N$ limit, though, the Polyakov line does not source glueballs and excited strings cannot decay, so the flux tube decouples from other bulk degrees of freedom.
Hence, at large $N$ the small $\tau$ asymptotics is to be reproduced entirely in terms of the exchange of highly-excited closed strings. This is the topic of section \ref{sec:IIA}.

In section \ref{sec:cardysec}, we then perform a Cardy trick of sorts, using the perturbative quark–antiquark potential to infer the large mass spectral density of closed string states contributing to the Polyakov correlator.\footnote{A similar open-closed duality analysis at the level of the EFT (large $\tau, R$) was performed by Lüscher and Weiss in \cite{Luscher:2002qv,Luscher:2004ib} (see also \cite{Aharony:2010cx}), and was recently generalized to the baryon junction in \cite{Komargodski:2024swh}.} For example, in \eqref{eq:D=4rho} we obtain that for  $D=4$
\begin{align}
	\rho_v( m,R) \sim \exp\left(\sqrt{\frac{\frac{24\pi}{11}R m}{\log   m}}\right)\,,
\end{align}
where $ m$ is the mass of the string, see (\ref{eq:D=3rho}) for $D=3$. Note that this is a milder growth than Hagedorn \cite{Hagedorn:1965st}, $\rho_H\sim \exp( m/T_H)$, as required from the absence of phase transitions at small $\tau$. This, in turn, implies a corresponding exponential decay for the coupling of closed flux tubes to the Polyakov line, see the discussion in section~\ref{sec:cardysec}.

In order to gain some insight on what type of worldsheet dynamics could lead to this high-energy behavior, in section \ref{sec:IV} we consider large $R$, and study the implications of the asymptotic scaling (\ref{eq:D=3rho}) for the Goldstone interactions, under the toy approximation of integrability. Using the Thermodynamic Bethe ansatz (TBA), we show that causality, in the form of positivity of time delays, implies a bound on the high-energy asymptotics of the form factor $K(p)\equiv\langle W|p,-p\rangle$ for the Wilson line to source pairs of Goldstones. Roughly, we find that
\begin{equation}
  \phantom{ at large p} 
  \qquad |K(p)|^2 \leq \frac{\lambda}{2p} \qquad 
  \text{ at large $p$}, 
\end{equation}
 with $\lambda$ the 't~Hooft coupling, defined here from the short-distance divergence of the correlator \eqref{eq:perturbativeasymp}, see (\ref{eq:decayK}) for the precise statement.
 Still in the integrable setting, in appendix \ref{app:zigzag} we show that the high-energy density (\ref{eq:D=3rho}) is incompatible with a linear phase shift $S\sim e^{ics}$ at high energies.
 This phase shift has been argued, via a semiclassical `zig-zag' model, to be the dominant high-energy scattering behavior \cite{Dubovsky:2018dlk,Dubovsky_2018}.

We conclude with non-integrable and bootstrappy speculations in section \ref{sec:V}. Several other appendices complement the main text.

\section{The Polyakov Correlator}\label{sec:secII}

\subsection{Cylinder partition function from large $N$}
\label{sec:IIA}

We consider the Euclidean correlation function of two Wilson loops (one fundamental and one antifundamental) in $\mathbb R^{D-1}\times S^1_R$,
\begin{equation}\label{eq:<WW>}
    \langle W_{\Box}(\tau,\vec 0) W_{\overline{\Box}}(0,\vec 0)\rangle\, \nonumber.
\end{equation}
The loops are parallel, separated by a Euclidean time $\tau$, and wrapping the compact direction of length $R$ (making them so-called Polyakov loops). We place them at the same position $\vec 0 \in \mathbb R^{D-2}$ in the transverse space. See figure \ref{fig:Cylinder}. Throughout, we keep the length of the compact direction above the inverse of the deconfinement temperature, $R>1/T_D$, so that the theory remains in the confining phase.

In this phase, fundamental Wilson loops source confining strings. At large $N$, the dominant contribution to the correlator comes from the exchange of single closed string states, whose internal dynamics we must sum over. In other words, the only worldsheet topology that survives the large $N$ limit is that of a cylinder stretching between the two loops, depicted in blue in figure \ref{fig:Cylinder}. This follows from a standard large~$N$ counting of diagrams \`a la 't~Hooft~\cite{tHooft:1973alw}. Higher-genus topologies, corresponding to the exchange of multi-string/glueball states, are suppressed by powers~of $1/N$.

The cylinder partition function $Z_{\text{cyl.}}(\tau ,R) \equiv   \langle W_{\Box} W_{\overline{\Box}}\rangle$ can be evaluated in two quantizations: 
\begin{equation}\label{eq:Zcyl}
    Z_{\text{cyl.}}(\tau ,R) = \langle W| e^{-H_{\text{cl.}} \tau}|W \rangle\ = \text{Tr}_{\mathcal H_\text{op.}}\hspace{-3pt}\left(e^{-H_{\text{op.}} R}\right) \,. 
\end{equation}
In the closed string channel, the Wilson line defines a state $|W\rangle$ in the Hilbert space $\mathcal H_\text{cl.}$ on $\mathbb R^{D-2}\times S^1_R$, which is evolved by the Hamiltonian $H_\text{cl.}$.\footnote{Equation \eqref{eq:Zcyl} imposes a Cardy-like condition \cite{Cardy:1989ir} on this state. A crucial difference with CFT, however, is that $Z_{\text{cyl.}}(\tau ,R)$ separately depends on $\tau/\ell_s$, $R/\ell_s$, rather than $\tau/R$, and so the two channels in \eqref{eq:Zcyl} are not related by a simple modular transformation.} At large $N$, $|W\rangle$ only couples to single closed string states which wind once around the $S^1_R$. In the open string channel, the correlator is interpreted as the thermal trace over the Hilbert space $\mathcal H_\text{op.}$ on $\mathbb R^{D-1}$ twisted by the Wilson line defects. At large $N$, the only states contributing to this trace are open strings stretching between the quarks.

Both representations should converge for any finite $\tau, R$ in the confining phase. 
While $R$ must be kept sufficiently large to remain in the confined phase,
we do not encounter any phase transitions as we dial $\tau$, since Wilson lines are mere probes of the theory. 
This is backed by both lattice simulations (see section \ref{sec:IIB}) and confining holographic models.\footnote{
In holography, Wilson lines define the endpoints of fundamental strings on the AdS boundary \cite{Maldacena:1998im}. 
While changing $R$ corresponds to tuning the length of the thermal circle, dialing $\tau$ just moves the endpoints of a string probing a fixed bulk geometry. The former induces a Hawking-Page phase transition at small $R$ \cite{Witten:1998zw}. In certain confining backgrounds, such as the Klebanov-Strassler \cite{Klebanov:2000hb}, the latter produces a quark-antiquark potential which explicitly interpolates between the linear and Coulomb regimes \cite{Cvicek:2007}.} Thus, $\tau$ provides a knob that allows us to smoothly interpolate between the UV and IR limits of the theory. 

\begin{figure}[t]
\centering
\includegraphics[width=0.8\linewidth]{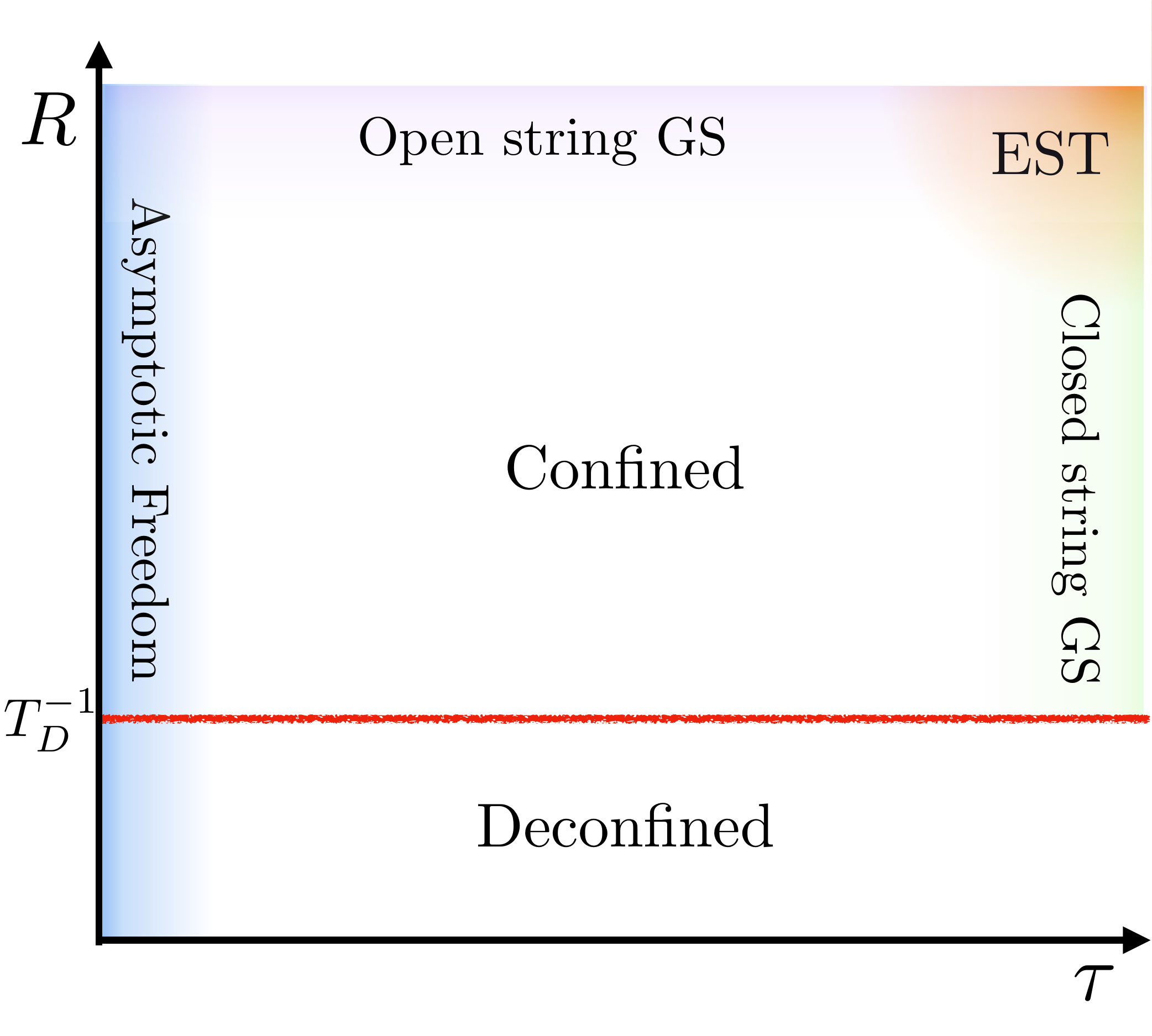}
\caption{Summary of the different limits of $Z_{\text{cyl.}}(\tau,R)$.}
\label{fig:correlator}
\end{figure}

We now discuss the dependence of the partition function  on the two dimensionless ratios $\tau/\ell_s$ and $R/\ell_s$, summarized in figure \ref{fig:correlator}. The decompositions (\ref{eq:Zcyl}) imply that 
\begin{align}
     \log Z_{\text{cyl.}}(\tau,R) &\xrightarrow[{\tau}/{\ell_s} \to \infty]{} -\tau E_0^\text{cl.}(R)\,, \label{{eq:larget}}\\
      \log Z_{\text{cyl.}}(\tau,R) &\xrightarrow[{R}/{\ell_s} \to \infty]{} -{R V_{q\bar{q} \label{{eq:largeR}}
      }( \tau)}\,,
\end{align}
where $E_0^\text{cl.}(R)$ is the ground state energy of the the winding flux tube, and $V_{q \bar{q}}(\tau)$ the energy of the open string ground state, identified here as the static quark-antiquark potential. 

When both $\tau/\ell_s$ and $R/\ell_s$ are large, the partition function can be evaluated in the long string EFT \eqref{eq:SEST}. The leading behavior is given by the area law
    \begin{equation}
         \log Z_{\text{cyl.}}(\tau,R) \xrightarrow[{\tau}/{\ell_s}, {R}/{\ell_s} \to \infty ]{} -\tau R /\ell_s^2  -2m_q R\,, \label{eq:largetR}
    \end{equation}
    which defines the scale $\ell_s^2$, and determines the leading behavior of $E_0^\text{cl.}(R), V_{q\bar{q}}(\tau)$ at large $R, \tau$ respectively. We also included the leading boundary effect, corresponding to an effective quark mass.\footnote{\label{foot:mq}The action (\ref{eq:SEST}) must be supplemented with boundary terms, see  \cite{Luscher:2002qv,Luscher:2004ib,Aharony:2010cx}, which we discuss in appendix \ref{app:EST}. In particular, these involve a boundary cosmological constant $m_q$, representing a quark mass, which is a counterterm that we can fix at will. Once $m_q$ is fixed, the constant piece in the short distance limit \eqref{eq:perturbativeasymp} is physical, and vice versa.}

 More important to us is that for $\tau \ll \ell_s$ 
 asymptotic freedom kicks in, and the leading asymptotics of the correlator $\langle W_{\Box}(\tau) W_{\overline{\Box}}(0)\rangle$ can be computed perturbatively; a computation which we now turn to.

\enlargethispage{-0.7in} 

\subsection{UV limit from asymptotic freedom}
\label{sec:IIB}

Indeed, 
at small $\tau/\ell_s$ with $R/\ell_s$ fixed we have (again!)
\begin{equation}
    \log Z_{\text{cyl.}}(\tau,R) \xrightarrow[{\tau/\ell_s \to 0 }]{} {-R V_{q\bar q}(\tau)}\,, \label{eq:finiteTasymp}
\end{equation}
where the quark anti-quark potential $ V_{q\bar q}$  is asymptotically Coulomb-like,
\begin{equation}
    V_{q\bar q}(\tau) = 
    \begin{cases}
    \frac{\lambda}{4\pi} \log(\tau)  + O(\tau^0)& \quad D=3,\\
   \frac{3 \pi}{11} \frac{1}{\tau\log(\tau)} + O\left(\frac{\log (\log(\tau))}{\tau \log(\tau)^2}\right)&\quad  D=4\,.
\end{cases}\label{eq:perturbativeasymp}
\end{equation}
The asymptotics in (\ref{eq:finiteTasymp}) is to be understood to the same order as (\ref{eq:perturbativeasymp}) at small $\tau/\ell_s$ fixed $R/\ell_s$. Subleading terms in (\ref{eq:finiteTasymp}) need not be extensive in $R$ and in general produce $R$-dependent corrections to $\log Z_{\text{cyl}}.$

Before discussing the origin of equations (\ref{eq:finiteTasymp}) and (\ref{eq:perturbativeasymp}), a few comments are in order.
First, the potential is only defined up to an additive constant.
The overall shift of the potential corresponds to a cosmological constant counterterm on the worldline of the defects, see footnote \ref{foot:mq}. 
Second, note the absence of a physical scale in the logarithms. This choice is deliberate and serves to emphasize that the scale is set at subleading orders in the short-distance expansion. 

In $D=4$, the leading asymptotics in (\ref{eq:perturbativeasymp}) is completely fixed with no undetermined parameters. This result is beautifully observed in the lattice, see for example the seminal work of Necco and Sommer \cite{Necco:2001xg} for the case of $SU(3)$.
In \eqref{eq:perturbativeasymp} we expressed the $D=4$ result directly at large~$N$.
In $D=3$, the functional form (\ref{eq:perturbativeasymp}) is under control but the $\lambda$ coefficient, which reduces to the 't~Hooft coupling $g^2N$ at large $N$, must be determined non-perturbatively in terms of the physical scale $\ell_s$. For $SU(3)$ it can be extracted from the data of Lüscher and Weisz \cite{Luscher:2002qv} to be $\lambda = 4.8 \ell_s^{-1}$, see also \cite{Brandt:2021kvt} for modern $SU(2)$ data. We are not aware of large $N$ determinations.

From the diagrammatic perspective, (\ref{eq:perturbativeasymp}) follows from the exponentiation of diagrams for the square Wilson loop \cite{Susskind:LesHouches1976,Fischler:1977yf,Appelquist:1977es,Gatheral:1983cz,Frenkel:1984pz}, combined with RG running of the coupling \cite{Gross:1973id,Politzer:1973fx}, which ensures that contributions to the potential beyond single-gluon exchange are subleading at small $\tau$.\footnote{One common point of confusion is that the free energy $\log Z_{\text{cyl.}}(\tau,R)$ depends both on the UV scale $\tau$ and the IR scale~$R$. The claim is that the leading singularity at small $\tau$ is generated by hard gluons with momenta $|k|\simeq1/\tau$ and, thus, it is perturbative. Contributions from gluons that propagate long distances and probe the scale $R$ or, more generally, non-perturbative effects, are less singular in the small $\tau$ expansion. This is, of course, backed by the lattice results which reproduce (\ref{eq:perturbativeasymp}).} For example, in $D=4$ we write
\begin{align}
    V_{q\bar{q}}(\tau) &= \frac{\lambda(\tau)}{8\pi \tau} \,,\nonumber\\
\tau\partial_\tau   \lambda(\tau) &= \frac{11}{24\pi^2} \lambda(\tau)^2 + O(\lambda(\tau)^3)\,, \label{eq:RG}
\end{align}
where in the RG equation we kept only the one-loop term, which is asymptotically correct at small $\tau$. Solving (\ref{eq:RG}) for small separations leads to (\ref{eq:perturbativeasymp}). The $D=3$ result in (\ref{eq:perturbativeasymp}) is manifest in perturbation theory.

So far we have discussed the potential which naturally dominates the limit $R/\ell_s \gg 1$. It turns out that (\ref{eq:perturbativeasymp}) also gives the correct small $\tau$ asymptotics at finite inverse temperature $R$, equation (\ref{eq:finiteTasymp}). This can be derived perturbatively, although the computation is more subtle \cite{Berwein:2017thy,Brown:1979ya,McLerran:1981pb,Nadkarni:1986cz,Nadkarni:1986as}. We comment on it in appendix \ref{app:Perturb}. When the dust settles, the small $\tau$ expansion is again determined from the exponentiation of diagrams, with leading asymptotics being controlled by hard gluons of momenta $|k| \simeq 1/\tau$ which are not affected by the finite thermal scale $R$.~The lack of temperature dependence in the short~distance limit of the quark anti-quark free energy (\ref{eq:finiteTasymp}) can also be observed (for finite $N$) directly from measurements of the Polyakov loop correlator in the lattice. See for example \cite{Bazavov:2008rw} for some data in the confined regime.\footnote{It would be great to have more comprehensive continuum extrapolated lattice data at short distances and finite but low temperatures for pure gauge theory, which seem to be lacking.} The asymptotics (\ref{eq:finiteTasymp})
continue to be valid above the deconfinement temperature, and there data is more plentiful \cite{Kaczmarek:2002mc, Weber:2017dmi}.

The remainder of this letter is devoted to exploring the implications of observation (\ref{eq:finiteTasymp}) for the worldsheet theory of confining strings.

\section{Imprints on the string spectrum}
\label{sec:cardysec}

In the closed string channel, the short-distance limit, $\tau/\ell_s\to 0$, of the cylinder partition function is controlled by the exchange of highly excited string states. We can thus extract the asymptotic spectral density by inverting the partition function. This logic mimics the seminal work by Cardy on the asymptotic density of states of two-dimensional CFTs \cite{Cardy:1986ie}. Rather than modular invariance, however, the asymptotic spectral density is controlled here by asymptotic freedom.

At large $N$, by inserting in \eqref{eq:Zcyl} a complete set of closed flux tube states $|n,\vec p_\bot\rangle \in \mathcal H_{\text{cl.}}$, labeled by their transverse center of mass momenta $\vec p_\bot$ and internal labels $n$, and summing over $\vec p_\bot$, we may write the partition function as \cite{Luscher:2004ib,Aharony:2010cx}
\begin{equation}\label{eq:ZfromK}
    Z_{\text{cyl.}}(\tau,R) = \int_0^\infty \hspace{-5pt} d m\, \rho_v( m,R)\, 2\tau \left(\frac{ m}{2\pi \tau}\right)^{\frac{D-1}{2}} \hspace{-5pt}K_{\frac{D-3}{2}}( m \tau)\,.
\end{equation}
Here,
\begin{equation}\label{eq:rho_B}
    \rho_v( m ,R) \equiv \sum_n |v_n(R)|^2 \delta\big( m - M_n^{\text{cl.}}(R)\big)\,,
\end{equation}
and $K_\nu(x)$ is the modified Bessel function of the second kind. The spectral density $\rho_v( m,R)$ counts the density of winding-one closed string states with mass $ m$, weighted by their coupling $v_n(R)\equiv \langle W_{\Box}|n,\vec p_\bot\rangle/\sqrt{2M^{\text{cl.}}_n(R)}$ to the Polyakov loops. Each such state contributes with a ${(D-1)}$-dimensional position-space propagator to the partition function \eqref{eq:ZfromK}. Details can be found in  appendix~\ref{app:rhos}.

By Poincar\'e invariance, adding a transverse separation $\vec x_\bot$ to the Polyakov loops only changes the partition function to $Z_{\text{cyl.}}\big(r\equiv\sqrt{\tau^2 + \vec x_\bot^2},R\big)$. The spectral density can be extracted from \eqref{eq:ZfromK} by passing to momentum space in the transverse direction, and performing an inverse Laplace transform in the center of mass frame,\footnote{\label{foot:muVSE}Applying directly an inverse Laplace transform would extract a density $\varrho(E,R)$ counting states by their (continuous) energy $E^{\text{cl.}}_{n,\vec p_\bot}(R) = \sqrt{M^{\text{cl.}}_n(R)^2 + \vec p_\bot^2}$, rather than their mass $M^{\text{cl.}}_n(R)$.
} i.e.
\begin{equation}\label{eq:rho-from-Z}
    \rho_v( m,R) = \frac{1}{2\pi i}\hspace{-3pt}\int_{\tau_0 - i\infty}^{\tau_0+i\infty} \hspace{-20pt}d\tau\, e^{ m \tau}\hspace{-3pt} \int d^{D-2}\vec x_\bot Z_{\text{cyl.}}\big(r,R\big)\,.
\end{equation}
Here $\tau_0 >0$, so that the contour runs to the right of all singularities of the integrand. This equation is checked explicitly in \eqref{eq:expl-inv-transf}.

In the large $ m$ limit, the integrals \eqref{eq:rho-from-Z} are dominated by the region where $r$ is small and $Z_{\text{cyl}}$ is controlled by (\ref{eq:finiteTasymp}) and (\ref{eq:perturbativeasymp}). Both the integral in $\vec x_\bot$ and the integral in $\tau$ can be evaluated by saddle point approximations. We obtain for the asymptotic density $\rho_v( m,R)$ as $ m \to \infty$:
\begin{itemize}
    \item[a.] \textit{In $D=3$,} where the leading inversion integrals can actually be done exactly, the result is
\begin{equation}
    \label{eq:D=3rho}
    \log\rho_v = \left({\frac{\lambda R}{4\pi}-2}\right) \log m
    + O( m^0),
\end{equation}
    with the error
stemming only from the corrections to the free energy \eqref{eq:finiteTasymp}. 

    \item[b.] \textit{In $D=4$,} 
\begin{align}\label{eq:D=4rho}
	\hspace{22pt} \log \rho_v = 2\sqrt{\frac{\frac{3\pi}{11}R m}{\log \sqrt  m}}
    + O\left(\frac{\sqrt{ m} \log(\log m)}{(\log  m)^\frac{3}{2}}\right)\,.
\end{align}
    Here the leading error has two sources; corrections to the free energy \eqref{eq:finiteTasymp} and corrections to the location of the saddle.
\end{itemize}

The detailed evaluation of the transforms and comments on subleading contributions can be found in appendices \ref{app:rhosD=3} and \ref{app:rhosD=4}.
As usual, the saddle approximation requires understanding the discrete density $\rho_\nu$ in an average sense, so that the methods are justified. This is especially important to subleading order. The results of the type (\ref{eq:D=3rho}) and (\ref{eq:D=4rho}) should be made sharp (perhaps with additional technical assumptions) by means of Tauberian results, following the CFT discussion \cite{Pappadopulo:2012jk,Das:2017vej,Qiao:2017xif,Mukhametzhanov:2018zja,Mukhametzhanov:2019pzy,Pal:2019zzr,Mukhametzhanov:2020swe,Das:2020uax}. 

In \cite{Diatlyk:2024qpr} and \cite{Kravchuk:2024qoh}, a similar strategy was followed to derive the asymptotic spectral density of the partition function of parallel defects in a CFT. Instead of perturbation theory, the small-$\tau$ behavior was fixed there by the effective theory for the fusion of conformal defects \cite{Diatlyk:2024zkk,Diatlyk:2024qpr,Kravchuk:2024qoh,Cuomo:2024psk}.
For one-dimensional defects, the partition function grows as $Z_{\text{CFT}}\sim e^{\frac{a_0 R}{\tau}}$, and the corresponding asymptotic density is $\rho_{\text{CFT}}(E) \sim E^{-\frac{3}{4}}\exp\left(2\sqrt{a_0 R E}\right)$, with $a_0$ the dimensionless Casimir energy. The $D=3$ density in \eqref{eq:D=3rho} differs significantly from this result because the Yang-Mills coupling is relevant and we have an explicit scale~$\lambda$. The $D=4$ result \eqref{eq:D=4rho}, on the other hand, only differs by log corrections because the Yang-Mills coupling is marginally relevant.

The spectral density $\rho_v( m,R)$ weights closed string states by their coupling $v_n(R)$ to the Polyakov loop. We can write it in terms of the density of states $\rho( m,R)$ as
\begin{equation}
    \rho_v( m,R) = \bar v( m,R) \rho( m,R)\,,
\end{equation}
where $\bar v( m,R) \equiv \frac{1}{N_ m}\sum_n^{N_ m} |v_n(R)|^2$ is the average of couplings over all the states with mass $ m$. Assuming a Hagedorn growth \cite{Hagedorn:1965st} for the closed string density of states, $\rho( m,R) \sim \exp( m/T_H)$, we conclude that the averaged coupling must decay exponentially $\bar v( m,R) \sim \exp(- m/T_H)$ as $ m\to \infty$ to recover the milder growths of (\ref{eq:D=3rho},\ref{eq:D=4rho}). This suppression reflects the absence of a Hagedorn-like phase transition as we vary $\tau$.\footnote{In comparison, for the critical bosonic string, the coupling of closed strings to D-branes is independent of the mass level and thus does not damp the Hagedorn growth. This leads to a divergence at finite $\tau$ associated to an open string tachyon. A similar divergence is found in type II superstrings stretching between a brane and an anti-brane \cite{Banks:1995ch}.

A separate mechanism for the absence of singularity in $\tau$ is available if the two boundaries of the cylinder are not conjugate to each other. In this case, states need not contribute positively to the spectral density, and the Hagedorn growth may be damped by virtue of cancellations between different states. This is the case for parallel D-branes in type II superstring \cite{Polchinski:1995mt}. In the main text, every flux tube state contributes positively.}
It would be interesting to verify directly this behavior in the lattice, and develop a microscopic understanding for it.

\section{Imprints on worldsheet dynamics}
\label{sec:IV}

In order to shed light into the microscopic description of the above result, we now explore what consequences asymptotic freedom might have for the dynamics on the string worldsheet. For simplicity, we will focus on the case of $D=3$ for the remainder of this letter. In lieu of a Lagrangian, we describe dynamics through on-shell scattering data of excitations on top of a very long winding string. We thus consider the cylinder partition at large $R$, where it is dominated by $V_{q\bar{q}}(\tau)$. While the dynamics of the bulk of the worldsheet is encapsulated in the S-matrices of the transverse Goldstone bosons $X^i$, the information of its boundary conditions is encoded in the corresponding R-matrices.

\begin{figure}[t]
\centering
  \includegraphics[width=0.8\linewidth]{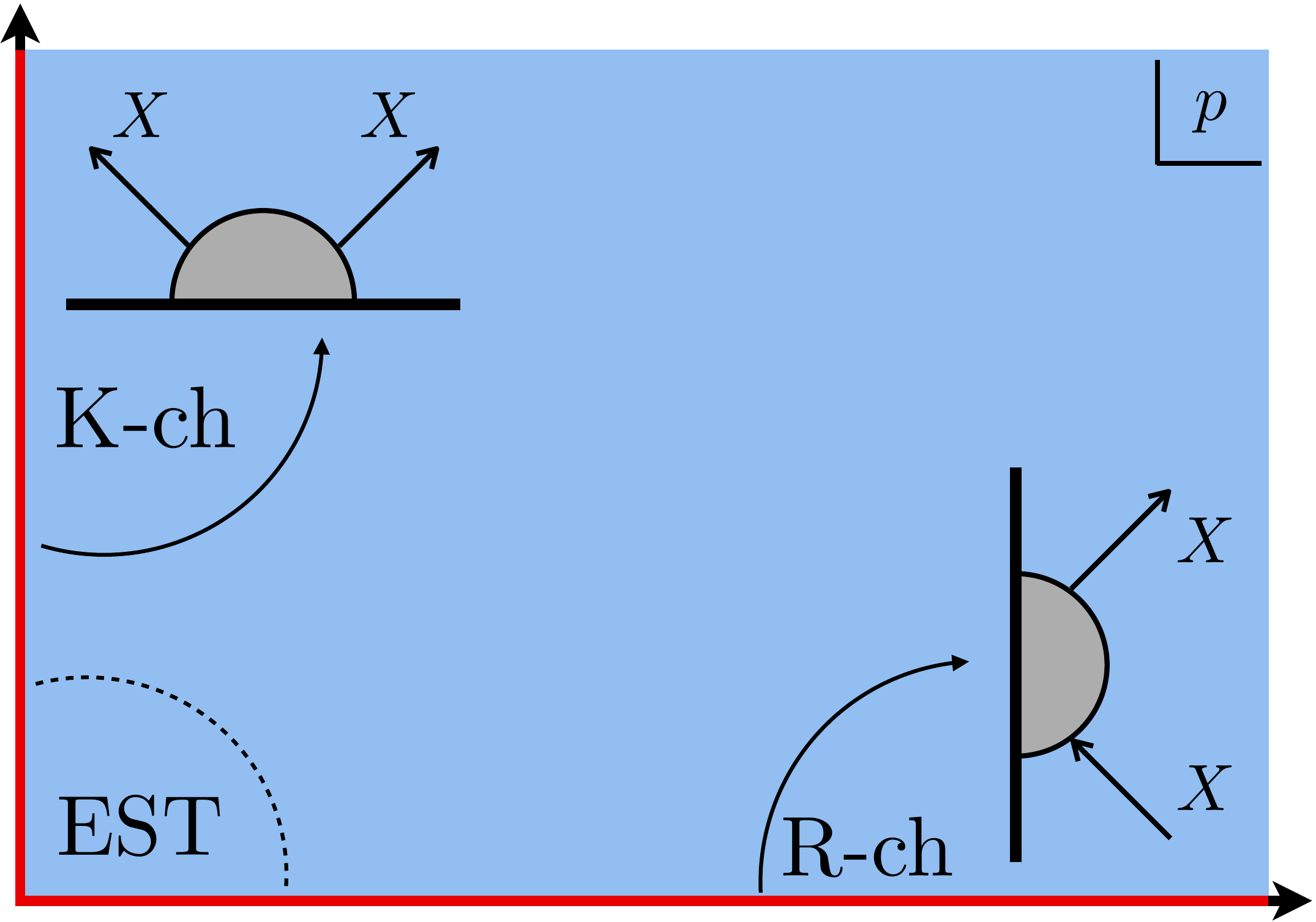}
\caption{The analytic structure of $R(p)$.}
\vspace{-1cm}
\label{fig:p-plane}
\end{figure}

In appendix \ref{app:R-matrix}, we offer a comprehensive review on reflection matrices.\footnote{We invite the reader to come back to this appendix, which can be read independently, where we also spell out the computation of the first terms of the R-matrix in the long string EFT, its result for the critical bosonic string, and some examples within integrability.} Here, we content ourselves by noting a few properties (summarized in figure \ref{fig:p-plane}) of the R-matrix $R(p)$ of one Goldstone boson scattering against the boundary. By causality, it is analytic in the upper-right complex $p$-plane. On the positive real axis, it has a cut from the physical exchange of states in the R-channel (in which the boundary runs along time), where it is bounded by unitarity; $|R(p)|\leq 1$. On the positive imaginary axis, it has another cut from states in the K-channel (in which the boundary runs along space). Here, unitarity imposes no constraints on the matrix $K(p) = R(ip)^*$.

In order to sharply connect the finite volume open string ground state energy $V_{q\bar{q}}(\tau)$ to the scattering data, we will assume integrability of the Goldstone interactions.
In this toy setup, we will be able to show that the result (\ref{eq:perturbativeasymp}) from asymptotic freedom bounds the asymptotic behavior of $K(p)$.
Of course, the worldsheet dynamics of pure Yang-Mills is definitely not integrable, and thus the toy models here are to be understood as proof of concept to motivate more general approaches, some of which we comment on in section \ref{sec:discussion}.

In the $D=3$ case, where the spectrum consists of a single Goldstone, the thermodynamic Bethe ansatz (TBA) \cite{Zamolodchikov:1989cf, LeClair:1995uf} expresses the open string ground state energy as\footnote{These formulas are easily derived from the standard torus TBA by treating $|K|^2$ as a chemical potential for the boundary to source pairs of particles. The linear term in (\ref{eq:E-eq}) is due to the cosmological constant of the $2d$ worldsheet theory, and we chose to tune the quark mass in \eqref{eq:largetR} away.
}
\begin{equation}\label{eq:E-eq}
    V_{q\bar q}(\tau) = \tau \ell_s^{-2} + \frac{1}{2\pi}\int_0^\infty dp\, \log \left(1-|K(p)|^2 e^{-\epsilon (p)}\right)\,,
\end{equation}
where the pseudoenergy $\epsilon(p)$ is the solution to
\begin{equation}\label{eq:eps-eq}
    \epsilon(p) = 2p \tau + \int_0^\infty \hspace{-6pt} \frac{dp'}{2\pi i} \,\frac{\partial \log S(pp')}{\,  \partial p'} \log\left(1-|K(p')|^2 e^{-\epsilon(p')}\right).
\end{equation}

The crucial observation is that the kernel of \eqref{eq:eps-eq} involves the Eisenbud-Wigner time delay $\Delta t = -i S^* \partial_p S$ \cite{eisenbud1948formal,Wigner:1955zz}. In relativistic massless scattering, causality requires that scattered particles come out inside the lightcones of the incoming ones, implying that $\Delta t\geq 0$ \cite{Adams:2006sv,Camanho:2014apa}. There has been some debate in the literature as to whether this is only up to a scale dictated by the uncertainty principle (see e.g.\ \cite{Chen:2023rar}). In appendix \ref{app:t-delay} we prove that, for $2d$ massless elastic S-matrices, $\Delta t\geq 0$ (for all energies) follows from unitarity and analyticity.

The absence of phase transitions as we dial $\tau$ implies that the $\log$ in (\ref{eq:E-eq},\ref{eq:eps-eq}) remains negative throughout its domain. This, combined with the positivity of the kernel in \eqref{eq:eps-eq}, readily implies a bound on the pseudoenergy; $\epsilon(p) \leq 2p \tau$. The intuition behind this bound is clear; positivity of the time delay makes interactions attractive, which in turn lower the energy of the states compared to the free theory.

By plugging this bound into \eqref{eq:E-eq} and using $\log(1-x)\leq-x$, we conclude that
\begin{equation}
    -V_{q\bar q}(\tau) \geq {-\tau \ell_s^{-2}+}\frac{1}{2\pi} \int_0^\infty dp \, |K(p)|^2 e^{-2p\tau}\,. \label{eq:causalitybound}
\end{equation}
We interpret this result as a causality bound on thermodynamic data. It states that the free energy can only be larger than its two-particle contribution. Since the logic behind this bound is not obviously tied to integrability, one might hope that it holds more generally. We comment on this possibility, and interesting bootstrap applications thereof in section \ref{sec:discussion}.

From (\ref{eq:causalitybound}) we have 
\begin{equation}
     -\frac{V_{q\bar q}(\tau) - \tau \ell_s^{-2}}{\log x}\geq \frac{e^{- 2 x \tau}}{2\pi \log x} \int_0^x dp \, |K(p)|^2,\nonumber
\end{equation}
where $x$ is large. Recalling that $V_{q\bar q}(\tau)$ asymptotes to the Coulomb potential \eqref{eq:perturbativeasymp} as $\tau\to 0$, we now take $x\rightarrow \infty$ with $\tau x $ fixed and then send $\tau x \rightarrow 0$. We obtain\footnote{Assuming the limit exists, otherwise we get a bound on the $\limsup$.}
\begin{equation}\label{eq:decayK}
    \lambda \geq \lim_{x\rightarrow\infty} \frac{2}{\log x}\int_0^x dp\, |K(p)|^2\,.
\end{equation}
This bounds $|K(p)|^2$ to decay no slower than $\lambda/2p$, on average, at large $p$.

Let us emphasize that $|K(p)|^2$ is not bounded by any fundamental principles of the scattering theory; this decay is a direct consequence of asymptotic freedom. Indeed, it rules out for example the asymptotic behavior of the R-matrix describing interactions on the boundary of a long critical string, discussed in appendix \ref{app:criticalST}. This asymptotic softness of the coupling between Goldstone bosons and the Wilson line is the microscopic reason for the decay of $\bar v( m,R)$ observed in the previous section. Of course, we have shown this in the restricted setting of integrability, but it seems plausible that a similar mechanism may persist beyond that.

We finish by noting that one can also use (\ref{eq:E-eq},\ref{eq:eps-eq}) to constrain the high-energy behavior of the S-matrix. Indeed, in appendix \ref{app:zigzag}, we show that an asymptotic behavior for $S(s)$ like that of a long Nambu-Goto string \cite{Dubovsky:2012sh} (also known as zig-zag asymptotics \cite{Dubovsky_2018}) given by $\log S \sim i c s$ is incompatible with the Coulomb asymptotics of $V_{q\bar{q}}(\tau)$ in (\ref{eq:perturbativeasymp}), regardless of the choice of $K$.

\section{Discussion}\label{sec:discussion}
\label{sec:V}

In this letter, we have initiated the systematic study of imprints that asymptotic freedom leaves on confining strings. By identifying a worldsheet observable which is controlled by perturbation theory, we have unlocked a door to connecting the flux tube theory with the underlying gauge theory. The asymptotic spectral densities from section~\ref{sec:cardysec}, and the bound on the integrable R-matrix from section~\ref{sec:IV} are but its first consequences. This investigation motivates many exciting research avenues, to which we look forward, and which we now discuss.

\vspace{-15pt}
\subsubsection*{Flux tube sum rules}
\vspace{-5pt}
Two-point functions of local operators $\langle \mathcal O(x) \mathcal O^\dagger(0)\rangle$ have played an undeniable role in the study of confining gauge theories \cite{Weinberg:1967kj,Shifman:1978bx,Shifman:1978by}. They probe the RG flow as we dial $x$ and, through various sum rules, they let us connect asymptotic freedom to the spectrum of mesons or glueballs that they couple to. As we have seen, the Polyakov correlator $\langle W_{\Box}(\tau) W_{\overline{\Box}}(0)\rangle$ probes the same flow, but it lets us explore instead the spectrum of flux tubes.
It turns out that, much like how one can write dispersive representations for $\langle \mathcal O \mathcal O^\dagger\rangle$, one can also do so for $\langle W_{\Box}W_{\overline{\Box}}\rangle$, which will be described elsewhere. It is then natural to ask whether the asymptotics \eqref{eq:finiteTasymp} can be used to derive sum rules on the low-lying flux tube spectrum. We hope to report back on this question soon.

An important tool in the study of the UV limit of local two-point functions is the OPE $\mathcal O_1(x) \mathcal O_2(0) \sim \sum_k c_k(x) \mathcal O_k(0)$. It predicts the structure of perturbative and non-perturbative corrections in the correlator. It would be very interesting to develop an analogous formalism for the fusion of Wilson lines, following the recent developments in the fusion of conformal defects \cite{Diatlyk:2024zkk,Diatlyk:2024qpr,Kravchuk:2024qoh,Cuomo:2024psk}. This should classify the structure of corrections in \eqref{eq:finiteTasymp}.
It is worth noting that in the phenomenology literature, this fusion is described in terms of pNRQCD \cite{Brambilla:1999xf,Brambilla:2004jw}, but its connection to conformal defect fusion is still lacking.

\vspace{-15pt}
\subsubsection*{Causality bounds on thermodynamics}
\vspace{-5pt}
One of the results of section \ref{sec:IV} is equation~\eqref{eq:causalitybound}. It is a bound on the finite-volume potential $V_{q\bar q}(\tau)$, thermodynamic data, stemming from causality. The logic for such a bound is the following. For a gas of particles (not necessarily dilute), thermodynamic potentials are built from the microscopic interactions of its constituents. If such interactions are relativistic, they are constrained by causality, which translates back into bounds for thermodynamic data. Of course, we only managed to make this sharp in the integrable setting, where TBA provides a clear link between finite-volume physics and scattering data, but the logic appears to transcend integrability.

The main obstruction to proving a bound such as \eqref{eq:causalitybound} beyond integrability is the lack of a sharp formula expressing thermodynamic potentials of a relativistic gas in terms of the S-matrices of its constituents. An early attempt was made by Dashen, Ma and Bernstein \cite{Dashen:1969ep,Dashen2}, who derived a formula for the free energy in $\mathbb R^{d-1}\times S^1_\beta$,
\begin{equation}\label{eq:DMB}
    \log Z(\beta) = \log Z_0(\beta) +\frac{1}{2\pi i}\int_0^\infty \hspace{-5pt} dE\, e^{-\beta E}\, \text{Tr}_c\left(\hat S^\dagger \partial_E \hat S\right)\,.
\end{equation}
Here $Z_0(\beta)$ is the partition function for the free gas, $\hat S(E)$ is an operator whose matrix elements in the Fock space are supposed to be on-shell S-matrices with center-of-mass energy $E$, and the subscript $c$ is a prescription to keep connected diagrams around the thermal circle.

While beautiful, this expression is highly formal. Indeed, the trace enforces that S-matrices be evaluated in the forward limit, and already the three-particle contribution
is highly ambiguous due to physical singularities \cite{Coleman:1965xm}. These issues have been recently revisited in \cite{Schubring:2024yfi,Baratella:2024sax,Baratella:2025fcj}, and the general consensus is that, to resolve this issue, S-matrices have to be taken off-shell. This challenges the statement that \eqref{eq:DMB} writes thermodynamics purely from scattering data. Hence, the DMB formula calls for a modern careful reformulation.\footnote{In particular, it is imperative to show that, for $2d$ integrable S-matrices, it (or an upgraded version thereof) reduces non-perturbatively to the expressions obtained from TBA.} Nevertheless, it is instructive to entertain its validity, and explore what could be learned from it.

The main observation is that, like in TBA, the kernel of \eqref{eq:DMB} can be interpreted as a time delay. Indeed, it is the so-called Wigner-Smith time delay operator \cite{Smith:1960zza,Martin:1976iw}, which applies to inelastic and multiparticle scattering. In this case, we still expect some notion of positivity from causality, but the precise statement is not as clear-cut. It would be outstanding to show that this kernel is positive, perhaps generalizing the methods from appendix~\ref{app:t-delay}. It would then follow that $Z(\beta)\geq Z_0(\beta)$.\footnote{Relations between thermodynamic inequalities like this one and EFT bounds have recently been explored in \cite{Fernandez-Sarmiento:2025tpy}.} Leveraging this to prove that \eqref{eq:causalitybound} and \eqref{eq:decayK} hold beyond integrability would require yet one more step; namely to develop a DMB formula for boundaries in terms of non-integrable R-matrices. These are all explorations that we hope to undertake in the future.

\vspace{-15pt}
\subsubsection*{R-matrix bootstrap}
\vspace{-5pt}
Finally, another aspect of the flux tube theory that deserves more attention is boundary Wilson coefficients of the low energy EFT. So far, these coefficients are measured from lattice simulations by fitting the open string ground state energy, see e.g.\ \cite{Brandt:2010bw,Billo:2012da,Brandt:2018fft,Brandt:2021kvt,Sharifian:2025fyl}. Now that the effective R-matrix is under control (see appendix \ref{app:EST}), it would be interesting to extend the method of \cite{Dubovsky:2013gi,Dubovsky:2014fma} to extract these coefficients from excited states, exploiting the low-energy integrability of the boundary.

Another approach would be to attempt to use bootstrap methods to constrain these coefficients, extending the logic of \cite{EliasMiro:2019kyf,EliasMiro:2021nul,Gaikwad:2023hof,Guerrieri:2024ckc} to R-matrices. One interesting target would be the sign of $b_2$, which is measured to be negative in lattice simulations of $SU(N)$ Yang-Mills theory \cite{Brandt:2018fft}. Could this sign follow directly from unitarity and other consistency conditions for $R(p)$? The answer is negative. Indeed, in appendix~\ref{app:CDD} we construct a healthy R-matrix that attains both signs of $b_2$ and, in fact, $b_2$ is already measured to be positive in the $3d$ Ising gauge model \cite{Billo:2012da,Caselle:2013dra}.

The power of the S-matrix bootstrap lies in the fact that $S(s)$ is bounded on the whole boundary of its domain of analyticity. The R-matrix, on the other hand, is unbounded on the imaginary axis, as discussed in appendix~\ref{app:R1-1} and \ref{app:FFunit}. This lets the Wilson coefficients of the boundary action take any value, preventing a systematic implementation of an R-matrix bootstrap program.\footnote{A notable exception is integrability. For integrable R-matrices, the relation \eqref{eq:integrable-R} allows to continue $R(p)$ to the full UHP, where it is bounded by unitarity on the whole real axis. This allows for an integrable R-matrix bootstrap study~\cite{Kruczenski:2020ujw}.} A bound on the sign of $b_2$ would thus need additional ingredients.
It would be very interesting if it were a consequence of asymptotic freedom. A strategy to test this could be to leverage \eqref{eq:decayK}, if proven beyond integrability, which provides an (asymptotic) bound for $R(p)$ on the imaginary axis.

{\begin{center}{\textbf{ACKNOWLEDGMENTS}} \end{center}}

It is a pleasure to thank Ofer Aharony, Michele Caselle, Clifford Cheung, Miguel Correia, Gabriel Cuomo, Sergei Dubovsky, Matthew Forslund, Victor Gorbenko, Igor Klebanov, Zohar Komargodski, Ryan Lanzetta, Juan Maldacena, Sebastian Mizera, Giuseppe Mussardo, Jo\~ao Penedones, Leonardo Rastelli, Rachel Rosen, Bruno Scheihing, David Simmons-Duffin, Joan Soto, Balt van Rees, Erez Urbach, Pedro Vieira, Xi Yin and especially David Gross for useful discussions.
We also thank Ofer Aharony, Igor Klebanov and Zohar Komargodski for comments on the draft.

This research was supported in part by grant NSF PHY-2309135 to the Kavli Institute for Theoretical Physics (KITP). This material is based upon work supported by the U.S. Department of Energy, Office of Science, Office of High Energy Physics, under Award Number DE-SC0011632.
This work was additionally supported in part by the Simons Foundation grant numbers 994312 and 917464 (Simons Collaboration on Confinement and QCD Strings).
JA is grateful to the KITP, the Aspen Center for Physics (supported by NSF grant no.~PHY-2210452), and ICTP-SAIFR (supported by FAPESP grant 2021/14335-0), where parts of this work were completed.

\pagebreak
\newpage
\bibliography{refs}


\appendix

\section{Perturbative computation}\label{app:Perturb}

 \enlargethispage{0.25in} 
 
In this appendix we briefly review the perturbative computation of the Polyakov loop correlator at small $\tau$ and finite $R$. We basically summarize the systematic computation of \cite{Berwein:2017thy} based on the technology of \cite{Gardi:2010rn,Gardi:2013ita}, and simply add a few comments for clarity and completeness. We note that the leading asymptotics (\ref{eq:finiteTasymp}) has been known since the '70s \cite{Brown:1979ya, McLerran:1981pb}. 

We are interested in computing, up to a $\tau$-independent normalization,
\begin{equation}
     \langle W_{\Box}(\tau) W_{\overline{\Box}}(0)\rangle\, = \delta_i^j\delta_l^k \langle L^i_j(\tau) \bar{L}^l_k(0)\rangle\,, \nonumber
\end{equation}
where $L^i_j$ ($\bar{L}^l_k$) is the untraced thermal Wilson line in the (anti-)fundamental, and we suppressed the transverse coordinate. It proves useful to decompose the correlator into the singlet and adjoint channels with respect to diagonal ($g(0) = g(\tau)$) gauge transformations,
\begin{equation}
    \delta_i^j\delta_l^k \langle L^i_j(\tau) \bar{L}^l_k(0)\rangle = \left(\frac{1}{N}\delta_i^k\delta_l^j  + 2 {T^a}_i^k {T^a}_l^j \right)\langle L^i_j(\tau) \bar{L}^l_k(0)\rangle. \label{eq:decomposition}
\end{equation}
Note that non-diagonal gauge transformations mix the singlet and adjoint contributions, and physical meaning should not be assigned to individual terms in (\ref{eq:decomposition}) except in the fusion limit $\tau \rightarrow 0$.\footnote{In this limit, the defects decompose into the direct sum of the identity operator and the Polyakov loop in the adjoint representation.
} This seems to be a common point of confusion in parts of the literature. In a fixed gauge, the separate contributions may also mix under renormalization. This separation is nevertheless convenient to organize the perturbative computation.

The diagrammatic expansion of individual terms in (\ref{eq:decomposition}) independently reexponentiates. Indeed, this is true at the level of the matrix of correlators $$\langle L^{i}_j \bar{L}^l_k\rangle:V^i(\tau)\otimes\overline{V}_k(0) \rightarrow V^j(\tau)\otimes \overline{V}_l(0),$$ henceforth denoted $\langle L \bar{L}\rangle$. The individual reexponentiation then follows since the color factor of individual diagrams preserves the decomposition into singlet and adjoint. This is straightforward to see in double-line notation which shows that for a diagram $D$ the color factor\footnote{Alternatively, one may note that $\langle L\bar L\rangle$ is an intertwiner with respect to diagonal gauge transformations, and invoke Schur's lemma to conclude that irreducible representations do not mix under compositions thereof.}
$$C(D)^{i l}_{j k} = c_1 \delta^i_j \delta^l_k + c_2 \delta^i_k \delta^l_j= \left(\tfrac{c_1 + N c_2}{N}\delta_i^k\delta_j^l  + 2 c_1 {T^a}_k^i {T^a}_j^l \right).$$
We thus write
\begin{equation}
    \langle W_{\Box} W_{\overline{\Box}}\rangle\, = e^{\tfrac{1}{N}\delta_i^k\delta_l^j (\log\langle L \bar{L}\rangle)^{i l}_{jk}}  + (N^2-1)e^{\tfrac{2{T^a}_i^k {T^a}_l^j}{N^2-1}(\log\langle L \bar{L}\rangle)^{i l}_{jk}}, \label{eq:correlatordecomp}
\end{equation}
and we are left with the task of computing the matrix logarithm $\log\langle L \bar{L}\rangle$, later to be projected into the singlet and adjoint channels.

The computation of $\log\langle L \bar{L}\rangle$ can be neatly performed by a replica-like trick \cite{Berwein:2017thy}. The idea is to compute $ \langle L\bar{L}\rangle^n$ analyticaly in $n$ and then evaluate
\begin{equation}
    \log \langle L \bar{L}\rangle  = \partial_n \langle L \bar{L}\rangle^n\vert_{n=0}. 
\end{equation}
The monomial $ \langle L \bar{L}\rangle^n$ can be computed diagrammatically. To do so, we draw a generic diagram $D$ in the presence of open lines $ L^{i}_j \bar{L}^l_k$ and assign a replica index $ \mu \in \{1,\dots,n\}$ to each disconnected cluster of internal particles (connectedness is with respect to internal lines).

The color factors of each cluster are then color ordered according to the replica index. The kinematic dependence $K(D)$ of the diagram factorizes. The overall contribution of the diagram to the monomial is obtained by summing over index assignments $\{\mu\}$,
\begin{equation}
     \langle L^{i}_j \bar{L}^l_k\rangle^n \supset_D K(D)\underbrace{\sum_{\{ \mu\}} C^{\{ \mu\}}(D)^{i l}_{jk}}_{C_n(D)}\,,
     \vspace{-15pt}
\end{equation}
 and we conclude
 \begin{equation}
     \log \langle L \bar{L}\rangle  =\sum_D K(D) \partial_n C_n(D)\vert_{n=0} .
 \end{equation}
 In sum, the logarithm is computed by summing over the complete set of diagrams but with modified color factors. The simplification comes from the fact that many diagram have vanishing color factors, see \cite{Gardi:2013ita} for a characterization of the non-vanishing diagrams.

Reference \cite{Berwein:2017thy} computes the projected logarithms of (\ref{eq:correlatordecomp}) to one-loop order in Coulomb gauge at small $\tau /R$ (with partial results at two-loops) in $D=4$. As expected, the result is as in (\ref{eq:finiteTasymp}). The leading contribution is given by the one-gluon exchange in the singlet channel, which matches the zero-temperature potential up to $\tau/R$ corrections. Higher-loop corrections are $\log(1/\tau)^{-1}$ suppressed once coupling running is taken into account. Adjoint contributions are exponentially suppressed and thus should be ignored in the perturbative small $\tau$ regime. As far as we know an analogous computation in $D=3$ has not been performed, but the overall logic should remain unchanged.

Finally, note that in naive perturbation theory the two exponentials in (\ref{eq:correlatordecomp}) can be expanded and the one-gluon exchange contribution cancels, the leading result coming from two-gluon exchange. This is clear since the color factor of a single gluon exchange in the Polyakov correlator vanishes. From this perspective, the small $\tau$ result at finite $R$ is quite subtle. The two-gluon exchange indeed gives the correct leading asymptotics at high temperature (i.e.\ $R\to 0$) \cite{Nadkarni:1986cz}, where the expansion of the exponentials in (\ref{eq:correlatordecomp}) is legal, but that is not justified at small $\tau$, fixed $R$, where they are exponentially large.

\vspace{-0.1 in}
\section{Closed string spectral densities}\label{app:rhos}
Consider the correlator \eqref{eq:<WW>} in the closed string channel, i.e.\ where the (Euclidean) time direction runs in the $\mathbb R^{D-1}$ factor. As discussed in the main text, the only states of the Hilbert space $\mathcal H_\text{cl.}\equiv\mathcal H(\mathbb R^{D-2}\times S^1_R)$ that contribute to this correlator in the large $N$ limit are single winding-one closed string states. We denote such states by $|n,\vec p_\bot\rangle$, where $n$ counts the oscillatory state of the string, and $\vec p_\bot$ measures its center of mass momentum in the transverse $\mathbb R^{D-2}$ directions. From a worldsheet perspective (in static gauge), $\vec p_\bot$ are (the conjugates of) the zero modes of the transverse $X^i$ Goldstone bosons.

The energy of such strings is given by a relativistic dispersion relation \cite{Meyer:2006qx}, $ 
    E^{\text{cl.}}_{n,\vec p_\bot}(R) = \sqrt{M^{\text{cl.}}_n(R)^2 + \vec p_\bot^2}\,,
$
where the string mass $M^{\text{cl.}}_n(R)$ depends only on the oscillatory level $n$ and the length $R$ of the compact direction.
In a Heisenberg picture, the Polyakov loop is moved in the time direction by $W_{\Box}(\tau,\vec 0) = e^{\tau \widehat H}W_{\Box}(0,\vec 0)e^{-\tau \widehat H}$, and in the transverse directions by conjugating with $e^{i\, x_\bot^j \cdot \widehat{P}_\bot^j}$. 

Inserting a complete basis of single-string states, and performing the integral over momenta \cite{Luscher:2004ib,Aharony:2010cx},
\begin{align}\label{eq:<WW>-app}
   & \langle W_{\Box}(\tau,\vec x_\bot)W_{\overline{\Box}}(0,\vec 0)\rangle =\\&\nonumber\, \sum_n \int \frac{d^{D-2}\vec p_\bot  e^{-\tau E^{\text{cl.}}_{n,\vec p_\bot}\hspace{-3pt}(R) + i\vec x_\bot \cdot \vec p_\bot}}{(2\pi)^{D-2} 2E^{\text{cl.}}_{n,\vec p_\bot}\hspace{-3pt}(R)}
    |\langle W_{\Box}(0,\vec 0)|n,\vec p_\bot\rangle|^2 \\
    =&\, \sum_n |v_n(R)|^2 \frac{\text{Vol}(S_{D-3})}{(2\pi)^{D-2}}\int_0^\infty \frac{|\vec p_\bot|^{D-3}d|\vec p_\bot| e^{-r E^{\text{cl.}}_{n,\vec p_\bot}\hspace{-3pt}(R) }}{ \sqrt{1+ |\vec p_\bot|^2/M^{\text{cl.}}_n(R)^2 }}
    \nonumber
    \\
    =&\, 
    \sum_n |v_n(R)|^2\, 2r
    \left(\frac{M_n^{\text{cl.}}(R)}{2\pi r}\right)^{\frac{D-1}{2}}
    K_{\frac{D-3}{2}}\left(r M_n^{\text{cl.}}(R)\right)\,.\nonumber
\end{align}
where we denote $r=\sqrt{\tau^2 + \vec x_\bot^2}$. Above we used the standard Lorentz invariant measure in $D-2$ dimensions, and we defined the overlap $|\langle W_{\Box}(0,\vec 0)|n,\vec p_\bot\rangle|^2\equiv 2M^{\text{cl.}}_n(R)|v_n(R)|^2 $ to match with the conventions in the literature. 
Note that $v_n(R)$ is independent of $\vec p_\bot$ due to boost invariance. 
In the second line, we introduced radial coordinates and simplified the integrand using Lorentz invariance. The position-space propagator comes about by changing coordinates to $y\equiv \sqrt{1+ |\vec p_\bot|^2/M^{\text{cl.}}_n(R)^2}$ and recalling the following integral representation of the Bessel function,
\begin{equation}
    K_\nu (z) = \frac{\sqrt{\pi} (\tfrac{1}{2}z)^\nu}{\Gamma(\nu + \tfrac{1}{2})} \int_1^\infty e^{-zy}(y^2-1)^{\nu - \frac{1}{2}}dy\,.
\end{equation}
In contrast with the main text, here we have chosen to add a transverse separation to the loops directly, for illustration purposes.

From a worldsheet perspective, \eqref{eq:<WW>-app} is the cylinder partition function $Z_{\text{cyl.}}(\sqrt{\tau^2 + \vec x_\bot^2},R)$, and $v_n(R)$ captures the overlap with the boundary state $|B\rangle$ generated by the action of the Wilson loop on the vacuum. Defining the mass spectral density $\rho_v( m, R)$ as in \eqref{eq:rho_B} readily reproduces \eqref{eq:ZfromK}. Conversely, the mass density can be extracted with the inverse transform in \eqref{eq:rho-from-Z}. This should be clear from the first expression in \eqref{eq:<WW>-app}, but it can be checked explicitly,
\begin{align}\label{eq:expl-inv-transf}
    &\int {d\tau\, d^{D-2}\vec x_\bot  d m}\, e^{ m' \tau} \rho_v( m,R)\, 2r \left(\frac{ m}{2\pi r}\right)^{\frac{D-1}{2}} K_{\frac{D-3}{2}}\left( m r\right) \nonumber \\
    &=\int \frac{d\tau\,  d m\, dy \, e^{ m' \tau} \rho_v( m,R)}{(2\pi)^{\frac{D-1}{2}}  \text{Vol}(S_{D-3})^{-1}}
   \frac{2 (y^2-1)^\frac{D-4}{2}}{\left( m\tau\right)^{\frac{1-D}{2}} y^{\frac{D-5}{2}}} K_{\frac{D-3}{2}}\left( m \tau y\right) \nonumber\\
    &=\int d\tau \,  d m\, e^{( m' -  m)\tau}
    \, \rho_v( m,R)\,= 
     2 \pi i \rho_v( m',R)\,,
\end{align}
where we changed to radial coordinates for $\vec x_\bot$, set $y\equiv\sqrt{1+\vec x_\bot^2/\tau^2}$, and performed the integrals. We have kept the integration domains implicit.

We now apply this transform to the UV result (\ref{eq:finiteTasymp},\ref{eq:perturbativeasymp}) in $D=3,4$, to extract the asymptotic spectral densities of closed string states contributing to the cylinder partition function in each case.

\subsection{\boldmath{$D=3$}}\label{app:rhosD=3}

In $D=3$, the $\tau\to 0$ limit of the partition function is $Z_{\text{cyl}}(\tau,R) \sim \tau^{-\frac{\lambda R}{4\pi}}$. The asymptotic spectral density is thus given by

\begin{align}
    \rho_v( m,R) \sim&\,\frac{1}{2\pi i}\hspace{-3pt}\int_{\tau_0 - i\infty}^{\tau_0+i\infty} \hspace{-20pt}d\tau\, e^{ m \tau}\int_{-\infty}^{\infty} dx_\bot \,(\tau^2 + x_\bot^2)^{-\frac{\lambda R}{8\pi}} \nonumber\\
    =&\,\frac{1}{2\pi i}\hspace{-3pt}\int_{\tau_0 - i\infty}^{\tau_0+i\infty} \hspace{-20pt}d\tau\, e^{ m \tau}\tau^{1-\frac{\lambda R}{4\pi}} 2\int_0^{\infty \, e^{-i \arg(\tau) }}\hspace{-35pt} du \,(1 + u^2)^{-\frac{\lambda R}{8\pi}}\,.
\end{align}

\noindent In the second line we used symmetry, and changed coordinates to $x_\bot = \tau u$. Note the $\arg(\tau)$ dependence of the contour. It can be freely rotated back to the real axis as long as $\frac{\lambda R}{4\pi} >1$, and the $u$-integral evaluates to a ratio of gamma functions. We are left with the inverse Laplace transform of a power, which is known to be $\mathcal L^{-1}\left\{\tau^{-a}\right\} =  m^{a-1}/\Gamma(a)$. 
Combining it all together, we find the asymptotic density
\begin{gather}\label{eq:rho-asymp-D=3}
    \rho_v( m,R) \sim \frac{\pi}{\Gamma\left(\tfrac{\lambda R}{8\pi}\right)^2}\left(\frac{ m}{2}\right)^{\frac{\lambda R}{4\pi}-2}\,.
\end{gather}

To quantify the corrections to this result, one must first control the corrections to \eqref{eq:finiteTasymp}. While we leave a careful analysis of such corrections for the future, on general grounds, we expect them to be of the form
$$\log Z_\text{cyl.} = -\frac{\lambda R}{4\pi} \log\tau + C(R) + O(\tau)\,,$$
where both the $\tau^0$ term and the $O(\tau)$ correction depend non-perturbatively on $R$.\footnote{The structure for perturbative corrections in $D=3$ is different from that in $D=4$ because $\lambda$ is dimensionful and it thus comes with powers of $\tau$ or $R$ attached.}
Carrying through the analysis above then leads to the following corrections for the asymptotic spectral density;
\begin{gather}\label{eq:rho-asymp-D=3}
    \rho_v( m,R) =  \pi \frac{e^{C(R)}}{\Gamma\left(\tfrac{\lambda R}{8\pi}\right)^2}\left(\frac{ m}{2}\right)^{\frac{\lambda R}{4\pi}-2}\Big(1+O(1/ m)\Big)\,.
\end{gather}
As usual, this is valid under sufficient smoothness assumptions so that the saddle analysis is justified. This is likely true only in a coarse-grained sense, see discussion in CFT where this is made sharp with Tauberian technology \cite{Pappadopulo:2012jk,Das:2017vej,Qiao:2017xif,Mukhametzhanov:2018zja,Mukhametzhanov:2019pzy,Pal:2019zzr,Mukhametzhanov:2020swe,Das:2020uax}.

\subsection{\boldmath{$D=4$}}\label{app:rhosD=4}
In $D=4$, the $\tau\to 0$ limit of the partition function is $Z_{\text{cyl}}(\tau,R) \sim \exp\big(-\frac{3\pi}{11}\frac{R}{\tau \log \tau} \big)$.
The corresponding~asymptotic spectral density is thus
\begin{align}\label{eq:D=4transform}
    \rho_v( m,R) \sim &\, \frac{1}{2\pi i}\hspace{-3pt}\int_{\tau_0 - i\infty}^{\tau_0+i\infty} \hspace{-25pt}d\tau\, e^{ m \tau}\hspace{-3pt} \int \hspace{-1pt} d^{2}\vec x_\bot 
    e^{-\frac{3\pi}{11}R/\sqrt{\tau^2+\vec x_\bot^2} \log \sqrt{\tau^2+\vec x_\bot^2}}\nonumber\\
     \sim&\, \frac{1}{i}\hspace{-3pt}\int_{\tau_0 - i\infty}^{\tau_0+i\infty} \hspace{-25pt}d\tau\, e^{ m \tau}
    \tau^2\int_0^{\infty \, e^{-i \arg(\tau) }} \hspace{-42pt} udu\,  \exp\left(-\tfrac{3\pi}{11}\tfrac{R}{\tau \log\tau}\tfrac{1}{\sqrt{1+u^2}}\right) \,,
\end{align}
where, again, in the second line we changed to radial variables and set $|\vec x_\bot|=\tau u$. Since the integral is dominated by small $\sqrt{\tau^2+\vec x_\bot^2}$, i.e.\ small $\tau$ and finite $u$, in the second line we have neglected $O(\tau(\log\tau)^2)^{-1}$ corrections to the exponent from expanding the log.

The $u$ integral can be evaluated via saddle point analysis. There is a single saddle at $u=0$, which the contour already steps on. We simply need to locally deform the contour around the saddle towards the real axis to ensure convergence of the Gaussian correction. To this order, the $u$-integral gives
\begin{equation}
	\int_0^{\infty \, e^{-i \arg(\tau) }} \hspace{-42pt} udu\,  \exp\left(-\tfrac{3\pi}{11}\tfrac{R}{\tau \log\tau}\tfrac{1}{\sqrt{1+u^2}}\right) = \frac{-\tau \log \tau}{\frac{3\pi}{11}R}e^{-\frac{3\pi}{11}\frac{R}{\tau \log\tau}}\,.
\end{equation}
Higher order (a.k.a. higher-loop) terms in the expansion around the saddle give additive corrections which do not modify the leading exponent, and are thus subleading.

The remaining $\tau$ integral can, in turn, also be evaluated by saddle point. The saddle equation reads
\begin{equation}
	\tau_*^2=-\frac{3\pi}{11}\frac{R}{ m}\frac{\log \tau_* + 1}{(\log\tau_*)^2}\,,
\end{equation}
which can be solved iteratively at large $ m$.
The leading solution is
\begin{equation}\label{eq:tau*}
	\tau_* = \sqrt{\frac{\frac{3\pi}{11}R}{ m \log \sqrt  m}} + O\left(\frac{\log(\log m)}{\sqrt{ m} (\log  m)^\frac{3}{2}}\right)\,.
\end{equation}
Shifting the contour so that it passes through the saddle, 
and performing the Gaussian integral, we find the result\footnote{The observant reader might notice that the overall power here differs from the CFT result \cite{Diatlyk:2024qpr,Kravchuk:2024qoh} quoted in section \ref{sec:cardysec}. This is because, with the transform in \eqref{eq:D=4transform}, we are computing a mass density rather than an energy density, as discussed in footnote \ref{foot:muVSE}.}
\begin{align}\label{eq:D=4rho-app}
	\rho_v( m,R) \sim&\, \sqrt{\pi}\left(\frac{\frac{3\pi}{11}R}{ m^3\log \sqrt m}\right)^\frac{3}{4} \exp\left(2\sqrt{\frac{\frac{3\pi}{11}R m}{\log \sqrt  m}}\right)\,.
\end{align}

There are various competing sources of error for this result. First, there are corrections to the asymptotic partition function \eqref{eq:finiteTasymp}, second, corrections from the consecutive saddle point approximations. For the former, while (again) we leave a careful treatment for the future, we expect
$$Z_{\text{cyl}}(\tau,R) = \exp\left(-\frac{3\pi}{11}\frac{R}{\tau \log \tau} + O\left(\frac{\log (\log\tau)}{\tau (\log\tau)^2}\right)\right)\,,$$
with the leading error coming from perturbative running.
Among the latter, the leading error comes from corrections to the location of the saddle \eqref{eq:tau*}. Amusingly, these two sources of error contribute at the same order, leading to
\begin{align}
	\rho_v( m,R) = \exp\left(2\sqrt{\frac{\frac{3\pi}{11}R m}{\log \sqrt  m}} + O\left(\frac{\sqrt{ m} \log(\log m)}{(\log  m)^\frac{3}{2}}\right)\right)\,,
\end{align}
where we have absorbed the prefactors from \eqref{eq:D=4rho-app} into subleading corrections of the exponent.

\section{Details of the R-matrix}\label{app:R-matrix}

Reflection matrices are often discussed in integrability following the seminal work of Ghoshal and Zamolodchikov \cite{Ghoshal:1993tm}, but they can be defined more generally. In this appendix, we review their definition and properties for general $2d$ theories with identical massless particles, henceforth referred to as Goldstones, as is needed for the $D=3$ flux tube. We also discuss some examples.

\subsection{General definition}
The R-matrices associated to a given boundary condition $B$ are defined from the overlaps between asymptotic states in the Hilbert space $\mathcal H_B$ on a semi-infinite line $\sigma^1\in [0,\infty)$ with boundary condition $B$ at $\sigma^1 = 0$,
\begin{equation}\label{eq:Rn-m}
    R_{n\to m} \equiv {}_{\hspace{5pt}B}^{\text{out}}\langle m | n\rangle^{\text{in}}_B = {}_{B_0}\langle m | \Omega_B(+\infty)^\dagger \Omega_B(-\infty) | n\rangle_{B_0}\,.
\end{equation}
This overlap describes the probability amplitude for $n$ left-moving states coming from infinity to reflect into $m$ right-moving states at $\sigma^1\to \infty$, see the bottom-right corner of figure \ref{fig:p-plane}. Here, the states $|n\rangle_{B_0}$ belong to the Fock space of an auxiliary free theory with ``free'' boundary conditions (which may be Neumann or Dirichlet), and $\Omega_B(t)\equiv e^{iH_Bt}e^{-iH_0t}$ is the analog of the usual M\o ller operator (see e.g.\ \cite{Weinberg:1995mt}) for the full Hamiltonian $H_B$ involving both bulk and boundary interactions.

A related object, the K-matrix, can be defined in the opposite quantization channel, i.e.\ where the boundary runs along the spatial direction. In this channel, the boundary defines a state $|B\rangle$ in the Hilbert space $\mathcal H$ on the infinite line $\sigma^1\in(-\infty,+\infty)$. The K-matrix is defined as the overlap between this state and the asymptotic states in $\mathcal H$,
\begin{equation}
    K_{B\to n} \equiv {}^{\text{out}}\langle n |B\rangle = \langle n | \Omega(+\infty)^\dagger|B\rangle\,. 
\end{equation}
This describes the probability amplitude for the boundary to emit $n$ particles into the bulk of the theory,\footnote{Absorption amplitudes, $K_{n\to B}$, can be defined analogously. They are related to emission amplitudes by $CPT$ invariance.} see the top-left corner of figure \ref{fig:p-plane}. R- and K-matrices are related by a double Wick rotation exchanging space and time. This can be regarded as a Cardy condition \cite{Cardy:1989ir} for the state $|B\rangle$ at the level of scattering data.

\subsection{Properties of the 1-to-1 matrix}\label{app:R1-1}
Let us now focus on the $1\to 1$ matrix, and discuss its properties. Factoring out the momentum-conservation delta functions, we write
\begin{subequations}
\begin{align}\label{eq:R1-1}
    R_{1\to 1} = &\,{}_{\hspace{5pt}B}^{\text{out}}\langle p | p'\rangle^{\text{in}}_B  \equiv 4\pi |p|\,  \delta(p+p') R(p)\,, \\ \label{eq:KB-2}
    K_{B\to 2} = &\,{}^{\text{out}}\langle p,p' | B\rangle  \equiv 4\pi |p|\,  \delta(p+p') K(p)\,,
\end{align}
\end{subequations}
where $R(p), K(p)$ are boundary values of analytic functions of the spatial momentum. Note that in the first line only energy is conserved, while in the second one it is the spatial momentum. This is due to the boundary breaking Lorentz invariance, which is in turn the reason why the functions depend on $p$ rather than a Mandelstam invariant.

The function $R(p)$ shares many features with a scattering amplitude $B X\to B X$ between a massless Goldstone and a putative particle $B$ of mass $m_B\to \infty$. At leading order in $m_B$, 2-momentum conservation is solved by
\begin{equation}
    p_1^ m = p_3^ m = (m_B,0), \quad p_2^ m = (p,-p), \quad p_4^ m = (p,p)\,,
\end{equation}
leading to the Mandelstam invariants $s=m_B^2 + 2m_B p$, $t=-4p^2$, $u=m_B^2-2m_B p$. We can use this analogy to infer the following properties of $R(p)$:

\paragraph{Analyticity.} Causality implies analyticity of the scattering amplitude for complex $s$ away from on-shell kinematics. This translates into analyticity of $R(p)$ in the first quadrant of the complex $p$-plane (see figure~\ref{fig:p-plane}). On the positive real axis, we hit the cut from $s\geq m_B^2$. On the positive imaginary axis, we hit the $t\geq 0$ cut.

\paragraph{Crossing.} The $s$-channel cut captures exchanges of physical intermediate states in the reflection amplitude. The crossed channel $BB\to XX$, on the other hand, reduces to the K-matrix, and the $t$-cut corresponds to on-shell Goldstones being emitted by the boundary. Crossing is the statement that one can analytically continue between the two channels; $R(ip)=K(p)^*$ for $p\in \mathbb R_{\geq 0}$.\footnote{The complex conjugate is due to approaching the $t$-channel cut from above. In a massive theory, the cuts open and one can access the physical side of the cut. See e.g.\ appendices A and E in \cite{Homrich:2019cbt}.}

\paragraph{Unitarity.} In the R-channel, unitarity is analogous to the familiar S-matrix case. Since the R-matrix operator $\hat R\equiv \Omega_B(+\infty)^\dagger \Omega_B(-\infty)$ in \eqref{eq:Rn-m} is unitary, inserting a complete basis of states in
${}_{\hspace{0.8pt}B_0} \langle p | \hat R^\dagger \hat R | p'\rangle_{B_0}$
shows that $|R(p)|^2 \leq 1$ there. In the K-channel, on the other hand, physical unitarity imposes no constraints. Indeed, from the $BB\to XX$ amplitude perspective, we sit on the generalized unitarity region where only massless intermediate states are allowed.
Boundedness kicks in above the physical threshold $t\geq 4m_B^2$, which has been pushed to infinity.

Before illustrating the above properties in some examples, let us mention that, given a valid R-matrix $R(p)$, one can always produce a new one $R_{\text{new}}(p)\equiv e^{i 2c p} R(p)$ by translating the boundary a distance $c$.\footnote{We thank Juan Maldacena for pointing this out to us.} Indeed, in the R-channel of figure \ref{fig:p-plane}, moving the boundary $B$ a distance $c$ to the right is equivalent to moving the wavepackets to the left with $|p\rangle_\text{new} = e^{-i \hat P c}|p\rangle$. Each particle thus pays a factor of $e^{icp}$. In the K-channel, the same is achieved by evolving the boundary state in Euclidean time; $|B\rangle_\text{new} = e^{-\hat H^\text{cl.} c}|B\rangle$. Acting it on the outgoing wavepackets produces a factor $K_\text{new}(p) = e^{-2pc}K(p)$, matching what we expect from crossing. As a result, the linear term in the phase of the R-matrix is ambiguous. This ambiguity drops out of propertly defined observables.\footnote{For example, in the TBA (\ref{eq:E-eq}), such a shift translates into a shift in $\tau$ and thus drops once the cylinder geometry is held fixed.}

\subsection{Example: Effective string theory}\label{app:EST}
As our main example, we consider the low-energy limit of the $D=3$ flux tube R-matrix.\footnote{The R-matrix in this theory is well-defined and escapes IR divergences in the same way as the S-matrix. Namely, by the absence of interactions between collinear Goldstones \cite{Dubovsky:2012wk}.}  Like the S-matrix, for ${p\ll 1/\ell_s}$, $R(p)$ is controlled by the effective theory of long strings. Apart from the effective action for the worldsheet~\eqref{eq:SEST}, it depends on the effective action for its boundary, $S_{\text{EST}}^{\text{bdy}}$, describing the low-energy limit of the Wilson line.
To leading order, these define Dirichlet boundary conditions, which receive higher-order corrections in EFT.   
The first few terms compatible with Poincar\'e invariance were classified in \cite{Aharony:2010cx};
\begin{equation}\label{eq:S-bdy}
    S_{\text{EST}}^{\text{bdy}} = \int_{\partial \Sigma} d\sigma^0 \Big(m_q + b_2\, \partial_0 \partial_1 X\cdot \partial_0 \partial_1 X + \cdots\Big)\,,
\end{equation}
where the constants $m_q, b_2$ are the first Wilson coefficients describing corrections on top of the Dirichlet boundary condition.

\begin{figure}[t]
\centering
\scalebox{1.3}{\begin{tikzpicture}
	\begin{pgfonlayer}{nodelayer}
		\node [style=none] (2) at (-3.35, -2.25) {};
		\node [style=none] (3) at (-3.35, 2.25) {};
		\node [style=none] (9) at (-0.65, -1.45) {};
		\node [style=none] (16) at (-0.65, 1.425) {};
		\node [style=none] (55) at (-3.35, 0) {};
		\node [style=none] (83) at (1.375, -2.25) {};
		\node [style=none] (84) at (1.375, 2.25) {};
		\node [style=none] (86) at (1.375, -1.15) {};
		\node [style=none] (89) at (3.075, -2.2) {};
		\node [style=none] (90) at (3.075, 2.175) {};
		\node [style=none] (98) at (1.375, 1.15) {};
		\node [style=none] (104) at (1.025, 1.15) {\scriptsize D};
		\node [style=none] (105) at (1.025, -1.15) {\scriptsize D};
		\node [style=none] (106) at (1.025, 0) {\scriptsize D};
		\node [style=none] (107) at (0, 0) {$+$};
		\node [style=none] (108) at (-1.6, 0) {};
		\node [style=none] (109) at (1.375, 0) {};
		\node [style=none] (110) at (3.125, 0) {};
		\node [style=none] (111) at (-1.725, -0.65) {\scriptsize NG};
		\node [style=none] (112) at (3.6, 0) {\scriptsize NG};
		\node [style=none] (113) at (-3.725, 0) {\scriptsize D};
	\end{pgfonlayer}
	\begin{pgfonlayer}{edgelayer}
		\draw [style=B] (2.center) to (3.center);
		\draw [style=B] (83.center) to (84.center);
		\draw (90.center) to (98.center);
		\draw (89.center) to (86.center);
		\draw (9.center) to (108.center);
		\draw [bend left=45] (55.center) to (108.center);
		\draw [bend right=45] (55.center) to (108.center);
		\draw (108.center) to (16.center);
		\draw [bend left=45] (109.center) to (110.center);
		\draw [bend right=45] (109.center) to (110.center);
		\draw [in=90, out=0, looseness=0.75] (98.center) to (110.center);
		\draw [in=-90, out=0, looseness=0.75] (86.center) to (110.center);
	\end{pgfonlayer}
\end{tikzpicture}}
\caption{Order $p^2$ contributions to the R-matrix.}
\label{fig:R-loop}
\end{figure}
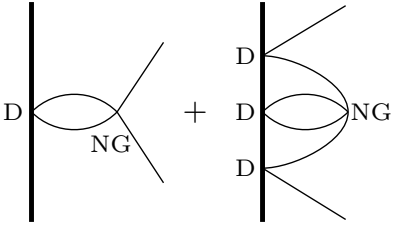
 
In what follows, we will compute the first terms of the reflection matrix separately in the $R$ and $K$ channels. We start by recalling the solution to the free equation of motion in the bulk of the string, $\partial^2 X=0$,
\begin{align}\label{eq:Xsol}
    X(\sigma) = \int_{0}^\infty &\frac{dp}{2\pi} \frac{\ell_s}{\sqrt{2p}} \bigg(a(p)e^{-ip(\sigma^0-\sigma^1)} + a^\dagger(p) e^{ip(\sigma^0-\sigma^1)}\nonumber\\
    &\quad + a(-p)e^{-ip(\sigma^0+\sigma^1)} + a^\dagger(-p) e^{ip(\sigma^0+\sigma^1)} \bigg)\,.
\end{align}
The operators $a^\dagger(p)$ with $p>0$ ($p<0$) create right-moving (left-moving) particles $|p\rangle \equiv \sqrt{2|p|} \, a^\dagger(p)|0\rangle$, while $a(p)$ annihilate them. They satisfy the canonical commutation relations
\begin{equation}
    [a(p), a^\dagger (q)] = 2\pi \delta(p-q)\,, \quad p\in \mathbb R\,,
\end{equation}
and so single-particle states are normalized by $\langle p| p'\rangle = 2|p| 2\pi \delta(p-p')$. The way the boundary condition enters this solution is different in each channel.

\subsubsection{R-channel}
For the R-channel computation, we take the boundary to extend along time at $\sigma^1=0$. We may consider two cases for the free boundary condition $B_0$:
\begin{itemize}
    \item Dirichlet: $\partial_0 X(\sigma^0, 0)=0$,
    \item Neumann: $\partial_1 X(\sigma^0, 0)=0$.
\end{itemize}
These conditions constrain the creation/annihilation operators by $a(-p) = \mp a(p)$, $a^\dagger(-p) = \mp a^\dagger(p)$ (upper signs for D and lower signs for N). Left- and right-moving states on the semi-infinite line are thus related by $|-p\rangle_{B_0} = \mp |p\rangle_{B_0}$. By computing the overlap \eqref{eq:R1-1}, we obtain $R(p) = \mp 1$.

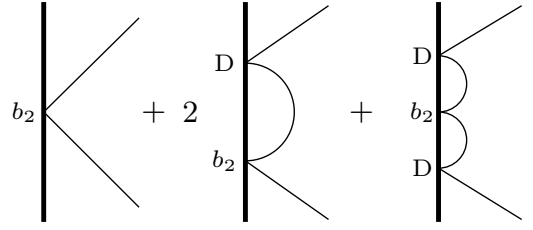
\begin{figure}[t]
\centering
\scalebox{1.3}{\begin{tikzpicture}
	\begin{pgfonlayer}{nodelayer}
		\node [style=none] (2) at (-4.5, -2.25) {};
		\node [style=none] (3) at (-4.5, 2.25) {};
		\node [style=none] (9) at (-2.55, -1.95) {};
		\node [style=none] (16) at (-2.55, 1.925) {};
		\node [style=none] (55) at (-4.5, 0) {};
		\node [style=none] (56) at (-0.375, -2.25) {};
		\node [style=none] (57) at (-0.375, 2.25) {};
		\node [style=none] (69) at (-0.375, -1) {};
		\node [style=none] (70) at (-0.375, 1) {};
		\node [style=none] (71) at (0.625, 0) {};
		\node [style=none] (73) at (1.325, -2.2) {};
		\node [style=none] (76) at (1.325, 2.175) {};
		\node [style=none] (83) at (3.575, -2.25) {};
		\node [style=none] (84) at (3.575, 2.25) {};
		\node [style=none] (86) at (3.575, -1.15) {};
		\node [style=none] (87) at (3.575, 0) {};
		\node [style=none] (88) at (4.15, -0.575) {};
		\node [style=none] (89) at (5.275, -2.2) {};
		\node [style=none] (90) at (5.275, 2.175) {};
		\node [style=none] (97) at (3.575, 0) {};
		\node [style=none] (98) at (3.575, 1.15) {};
		\node [style=none] (99) at (4.15, 0.575) {};
		\node [style=none] (100) at (-2.25, 0) {$+$};
		\node [style=none] (101) at (2, 0) {$+$};
		\node [style=none] (102) at (-0.8, 1) {\scriptsize D};
		\node [style=none] (103) at (-0.8, -1) {\scriptsize $b_2$};
		\node [style=none] (104) at (3.25, 1.15) {\scriptsize D};
		\node [style=none] (105) at (3.25, -1.15) {\scriptsize D};
		\node [style=none] (106) at (3.25, 0) {\scriptsize $b_2$};
		\node [style=none] (107) at (-1.5, 0) {$2$};
		\node [style=none] (108) at (-4.9, 0) {\scriptsize $b_2$};
	\end{pgfonlayer}
	\begin{pgfonlayer}{edgelayer}
		\draw [style=B] (2.center) to (3.center);
		\draw [style=B] (56.center) to (57.center);
		\draw [bend left=45] (70.center) to (71.center);
		\draw [bend left=45] (71.center) to (69.center);
		\draw [style=B] (83.center) to (84.center);
		\draw [bend left=45] (87.center) to (88.center);
		\draw [bend left=45] (88.center) to (86.center);
		\draw [bend left=45] (98.center) to (99.center);
		\draw [bend left=45] (99.center) to (97.center);
		\draw (55.center) to (9.center);
		\draw (16.center) to (55.center);
		\draw (76.center) to (70.center);
		\draw (73.center) to (69.center);
		\draw (90.center) to (98.center);
		\draw (89.center) to (86.center);
	\end{pgfonlayer}
\end{tikzpicture}}
\caption{Order $p^3$ contributions to the R-matrix.}
\label{fig:R-tree}
\end{figure}

Interactions are introduced in complete analogy with standard scattering theory. Namely, we take the R-matrix operator
\begin{equation}
    \hat R = T\left\{e^{i\int d^2 \sigma\, \mathcal L_\text{int}^\text{bulk} + i \int d\sigma^0 \mathcal L_\text{int}^\text{bdy}}\right\}\,,
\end{equation}
and expand it in perturbation theory. For the $D=3$ flux tube ending on a Wilson line, the leading bulk and boundary interactions are
\begin{align}
    \mathcal L_\text{int}^\text{bulk} =&\, \frac{1}{8\ell_s^2} (\partial_a X \partial^a X)^2 + \cdots\,,\\
    \mathcal L_\text{int}^\text{bdy} =&\, b_2\, (\partial_0 \partial_1 X)^2 + \cdots\,,
\end{align}
and we are instructed to take Dirichlet boundary conditions for $B_0$. The contributions up to order $\sim p^3$ are depicted in figures \ref{fig:R-loop} and \ref{fig:R-tree}. The result is\footnote{Here we quote the finite contributions. There is also an $O(p)$ piece with diverging coefficient that can be canceled by tuning the linear phase from moving the boundary.} 
\begin{equation}\label{eq:REST}
    R(p) = - \exp i\left(\frac{\ell_s^2}{2}p^2 + 4b_2 \ell_s^2 p^3 +\cdots  \right)\,.
\end{equation}

\subsubsection{K-channel}
For the K-channel computation, we take the boundary to extend along space at $\sigma^0=0$. The free boundary state $|B_0\rangle$ is obtained by acting the boundary conditions on it:
\begin{itemize}
    \item Dirichlet: $\partial_1 X(0, \sigma^1) |B_0\rangle =0$,
    \item Neumann: $\partial_0 X(0, \sigma^1) |B_0\rangle =0$.
\end{itemize}
In terms of creation/annihilation operators, these respectively imply $\left(a(p) \pm a^\dagger(-p)\right)|B_0\rangle = 0$ ($p\in \mathbb R$). It is easy to see that these equations are solved by
\begin{equation}
    |B_0\rangle = \mathcal N \exp\left(\mp\int_0^\infty \frac{dp}{2\pi} a^\dagger(p) a^\dagger(-p)\right)|0\rangle\,,
\end{equation}
since~${[a(p), e^{\gamma \int \frac{dq}{2\pi} a^\dagger(q) a^\dagger(-q)}] = \gamma e^{\gamma \int \frac{dq}{2\pi} a^\dagger(q) a^\dagger(-q)} a^\dagger(-p)}$. By taking the overlap \eqref{eq:KB-2}, we obtain $K(p) = \mp \mathcal N$. Crossing (equivalently, the Cardy condition) fixes the normalization of the state to $\mathcal N=1$.

Interactions in the K-channel are introduced in two steps. First, we dress the free boundary state with the boundary action,
\begin{equation}
    |B\rangle = e^{S_{\text{EST}}^\text{bdy}(\sigma^0=0)}|B_0\rangle\,,
\end{equation}
which we treat perturbatively. Second, we evolve to asymptotic states with the bulk interaction Lagrangian,
\begin{equation}
    K_{B\to 2} = \langle p,p'|T\big\{e^{i\int d^2 \sigma\, \mathcal L_\text{int}^\text{bulk}}\big\}|B\rangle\,.
\end{equation}
Up to order $\sim p^3$, the diagrams contributing are again those in figures \ref{fig:R-loop} and \ref{fig:R-tree} (rotated by 90 degrees), giving
\begin{equation}
    K(p) = -\exp\left( i \frac{\ell_s^2}{2}p^2 + 4b_2 \ell_s^2 p^3 + \cdots\right)\,.
\end{equation}
Comparing to \eqref{eq:REST}, we can verify explicitly the crossing equation $R(ip)=K(p)^*$.\footnote{The overall coefficient in $R$ and $K$ can be fixed by tuning the cosmological constant $m_q$ in \eqref{eq:S-bdy}. Here we have chosen to turn it~off.}

\subsubsection{Low-energy integrability}
It is worth pointing out that, up to the order quoted, \eqref{eq:REST} is integrable. The R-matrices for integrable boundaries of integrable theories satisfy additional conditions beyond those listed in section \ref{app:R1-1}. These were spelled out in \cite{Ghoshal:1993tm,Caselle:2013dra}, and for flavorless scattering they read
\begin{equation}\label{eq:integrable-R}
    R(p) R(-p) = S(p,p)\,, \quad |R(p)|^2 =1 \quad (p>0).
\end{equation}
Indeed, recalling the $D=3$ EFT expansion for the S-matrix \cite{Dubovsky:2012sh,EliasMiro:2019kyf}, $S(p,p') = \exp(i \ell_s^2 p p'+O(pp')^3)$, we immediately see that \eqref{eq:REST} satisfies these additional conditions. This highlights that the low-energy integrability known for the flux tube bulk extends also to its boundaries. It remains an important open question to determine at which order in $p$ boundary integrability is lost.

The low-energy integrability of the S-matrix has been used extensively to fit bulk Wilson coefficients against closed string lattice data \cite{Dubovsky:2013gi,Dubovsky:2014fma}. It would be interesting to extend that analysis to boundary Wilson coefficients. An early step in this direction was taken in \cite{Caselle:2013dra}, where, by computing the open string ground state energy using the TBA equations (\ref{eq:E-eq},\ref{eq:eps-eq}) and comparing to the results of \cite{Aharony:2010db}, they fixed the R-matrix to \eqref{eq:REST} (up to an overall sign, which does not enter the TBA equations). Higher boundary Wilson coefficients remain largely unexplored.

\subsection{Example: Critical bosonic string}\label{app:criticalST}
As our second example, we briefly discuss the bosonic string in critical dimension ($D=26$). The S-matrix is known to be reflectionless and given by the leading Nambu-Goto phase shift \cite{Dubovsky:2012wk},
\begin{equation}\label{eq:S-critical}
    S_{ab}^{cd}(p,p') = e^{i\ell_s^2 pp'}\delta_a^c \delta_b^d\,,
\end{equation}
where $a,...,d$ run over the $D-2$ transverse directions. The string can end on D-branes. The R-matrices of the corresponding boundaries are diagonal and given by
\begin{equation}\label{eq:R-critical}
    R_a^b(p) = \mp e^{i\ell_s^2 p^2/2} \delta_a^b\,,
\end{equation}
with the upper sign for Dirichlet directions and the lower one for Neumann directions.

The above result for the R-matrix should follow in a diagrammatic computation 
from cancellations in the critical dimension, see for example \cite{Dubovsky:2012sh} for the cancellation of the Polchinski-Strominger term \cite{Polchinski:1991ax} in the $D=26$ $S$-matrix. Here, following \cite{Dubovsky:2012wk}, we bypass this computation by exploiting integrability and comparing the ground state energy obtained from TBA to the known result. The general TBA equations for particles with flavor are very intricate. For diagonal K- and S-matrices ($K^a, S_{ab}$), however, they are straightforward generalizations of (\ref{eq:E-eq},\ref{eq:eps-eq}):
\begin{align}
    E_0^{\text{op.}}(\tau) =&\, \tau \ell_s^{-2} + \frac{1}{2\pi}\sum_a\int_0^\infty dp\, L_a(p)\,,\\
    \epsilon_a(p) =&\, 2p \tau 
    +\sum_b \int_0^\infty dp'\, \frac{\partial \log S_{ab}(p,p')}{2\pi i\,  \partial p'} L_b(p')\,.\nonumber
\end{align}
Here, $L_a(p) \equiv \log \left(1-K_{B_1}^a(p)K_{B_2}^a(p)^* e^{-\epsilon_a (p)}\right)$, where we are allowing for the left and right boundaries to be different.

These equations are easily solved for the S- and R-matrices above, giving
\begin{equation}
    E_0^{\text{op.}}(\tau) = \sqrt{\frac{\tau^2}{\ell_s^4} - \frac{2\pi}{\ell_s^2}\left(\frac{D-2}{24} -\frac{3n_{\text{DN}}}{48}\right)}\,,
\end{equation}
where $n_{\text{DN}}$ is the number of transverse directions with Neumann boundary conditions on one end and Dirichelt on the other. This matches the long-known ground state energy of the critical bosonic open string \cite{Siegel:1976qn,Arvis:1983fp}. Indeed, we recognize the Casimir energies of the D-D/N-N directions and the D-N directions. We thus conclude that \eqref{eq:S-critical} and \eqref{eq:R-critical} do describe the critical bosonic string and its boundaries.\footnote{For $D\neq 26$, they still describe a valid $2d$ theory, albeit one that is not compatible with spacetime Poincar\'e invariance.}

\subsection{Example: CDD factors}\label{app:CDD}
Our final example is more academic. The flavorless integrability conditions for a massless S-matrix are $S(s)^*=S(-s^*)$ and $|S(s)|^2=1$ (for $s>0$). A general meromorphic solution to these equations can be written as \cite{Zamolodchikov:1991vx} (see also appendix \ref{app:t-delay})
\begin{equation}\label{eq:SCDD}
    S_{\text{CDD}}(s) = \prod_j \frac{im_j^2 - s}{im_j^2 + s}e^{i 2\delta_0(s)}\,,
\end{equation}
where $\delta_0(s)$ is analytic in the UHP. The masses $m_j$ are known as CDD zeros (or resonances) \cite{Castillejo:1955ed}, and they must be real or come in pairs $|m_j|e^{\pm i \alpha}$ with $\alpha\in (0,\pi/4)$. The corresponding R-matrices must satisfy \eqref{eq:integrable-R}. Its minimal solution is
\begin{equation}
    R_\text{CDD}(p) = \prod_j \frac{e^{i\frac{\pi}{4}} m_j - 2p}{e^{i\frac{\pi}{4}} m_j - i2p} e^{i 2\delta_0(4p^2)/2}
    \,.
\end{equation}
More solutions can be obtained by multiplying by solutions of $R(p)R(-p)=1$. Solutions!

Systematic S-matrix bootstrap bounds tend to be saturated by integrable solutions \cite{Paulos:2016but,EliasMiro:2019kyf}. While there is no general boundary counterpart to the bootstrap,
as discussed in section~\ref{sec:discussion},
we can still explore the space of theories ruled in by the above CDD factors. This unveils at least a chunk of the space of Wilson coefficients allowed by analyticity, crossing and unitarity. Take, for example, $R_\text{CDD}(p)$ with one pair of complex CDD zeros $m_j = m e^{\pm i \alpha}$. Expanding the phase shift at low energies and comparing it with \eqref{eq:REST}, we obtain
\begin{equation}
    m^2\ell_s^2 = 16\cos 2\alpha\,, \qquad m b_2 = \frac{1}{6\sqrt{2}}\frac{\cos 3\alpha}{\cos 2\alpha}\,.
\end{equation}
In particular, we note that by dialing $\alpha\in (0,\pi/4)$ we can attain any sign for $b_2$.

\subsection{Boundary form factor bootstrap?}\label{app:FFunit}
We close this appendix by revisiting the unitarity conditions on the R-matrix, discussed in subsection \ref{app:R1-1}. We argued that $|K(p)|^2$ is not bounded by unitarity. As discussed in section \ref{sec:discussion}, this poses an obstruction to the implementation of a systematic R-matrix bootstrap program. The K-matrix ${}^{\text{out}}\langle p,p' | B\rangle$ looks strikingly similar to a local-operator form factor ${}^{\text{out}}\langle p,p' | \mathcal O(x)\rangle$. For the latter, a systematic bootstrap program was initiated in \cite{Karateev:2019ymz}, so one might wonder whether a similar strategy would work for boundaries. Here we sketch why that is not the case.

The strategy to mimic form-factor unitarity would be the following. We start by constructing the operator $|\psi(E)\rangle \equiv \int dt\, e^{i (E-H^\text{cl.}) t}|B\rangle$, whose inner product with itself produces a spectral density,
\begin{equation}
    \langle \psi(E')|\psi(E)\rangle =
     4\pi^2 \delta(E-E') \varrho(E,R)\,,
\end{equation}
where $\varrho(E, R) \equiv \sum_n |\langle n |B\rangle|^2\delta(E-E_n)$.\footnote{In the context of the string worldsheet, this density would be the one discussed in footnote \ref{foot:muVSE}, counting states by their total energy rather than their mass.} Here we are regulating the theory by putting it on a large but finite ring of length $R$ (hence why a sketch), and we inserted a complete basis of states $|n\rangle$ on the $S^1_R$ Hilbert space.

Next, we construct a matrix of inner products with the states $|p_1,p_2\rangle^\text{out}$ and $|\psi(E)\rangle \equiv \int dt\, e^{i (E-H^\text{cl.}) t}|B\rangle$,
\begin{equation}\label{eq:FFmatrix}
   \begin{pmatrix}
       {}^\text{out}\langle p_1'p_2'|p_1p_2\rangle^\text{out}
       & {}^\text{out}\langle p_1'p_2'|\psi(p_1-p_2)\rangle  \\
       \langle \psi(p_1'-p_2')|p_1p_2\rangle^\text{out}
       & \langle \psi(p_1'-p_2')|\psi(p_1-p_2)\rangle
   \end{pmatrix}\,.
\end{equation}
In the first entry, we have the the kinematic delta function $4\pi^2 4|p_1p_2| \delta(p_1-p_1') \delta(p_2-p_2')$, in the off-diagonal entries; the K-matrices $8\pi^2|p_1'|\delta(p_1'+p_2')\delta(E-p_1'+p_2') K(p_1')$ and $8\pi^2|p_1|\delta(p_1+p_2)\delta(E'-p_1+p_2) K(p_1)^*$, and the spectral density in the last entry.

The matrix \eqref{eq:FFmatrix} must be positive semidefinite by unitarity. In particular, its determinant,
\begin{align}
    &\frac{32\pi^3R}{|p_1p_2|^{-1}}\delta(p_1-p_1')\delta(p_2-p_2')\left(\varrho(2p_1,R) -\frac{R}{4\pi}|K(p_1)|^2\right)\,,\nonumber
\end{align}
must be non-negative. This leads to the bound $4\pi \varrho(2p, R) \geq R |K(p)|^2$. By taking a Laplace transform in the energy $E=2p$, we can write the bound in terms of the cylinder partition function,
\begin{equation}\label{eq:Z>K}
   Z_{\text{cyl.}}(\tau ,R) \geq \frac{R}{2\pi} \int_0^\infty dp\, |K(p)|^2 e^{-2p\tau}.
\end{equation}

This appears to provide an upper bound for $K$ in terms of $Z_{\text{cyl.}}(\tau ,R)$, but one should not forget that asymptotic states are defined in infinite volume. As $R\to \infty$, the left hand side is dominated by the ground state energy, $Z_{\text{cyl.}}\to e^{-RE_0^\text{op.}(\tau)}$. We are working here in the standard convention in scattering theory, where the ground state energy $E_0^\text{op.}(\tau)$ vanishes in the infinite volume limit $\tau \to\infty$. Monotonicity of $Z_{\text{cyl.}}$ then implies that,\footnote{$Z_{\text{cyl.}}(\tau,R)$ is monotonically decreasing in each of its arguments (in fact, it is a completely monotone function). This is clear from its Hamiltonian formulation in each channel, e.g.\ $Z_{\text{cyl.}} = \langle B|e^{-\tau H^\text{cl.}}|B\rangle$, and the fact that $H^\text{cl.}$ is bounded below.} for finite~$\tau$, $E_0^\text{op.}(\tau)\leq 0$. Therefore, the left hand side in \eqref{eq:Z>K} diverges and the bound trivializes.

The bound \eqref{eq:causalitybound} that we derived in the main text is qualitatively similar to \eqref{eq:Z>K}, but it features the free energy rather than the partition function. Then, both sides of the inequality are extensive, and the bound remains in the infinite volume limit.

We finish by noting that an argument analogous to the one sketched here does work when considering correlators in finite volume rather than asymptotic states. See \cite{Meineri:2025qwz} for applications to the bootstrap of conformal boundaries in $2d$ CFTs.

\vspace{-0.2 in}
\section{Positivity of time delays}\label{app:t-delay}
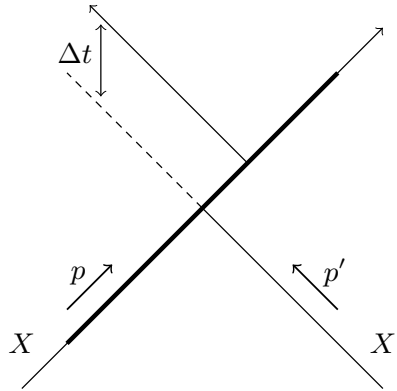
\begin{figure}[t]
\centering
\scalebox{1.2}{\begin{tikzpicture}
	\begin{pgfonlayer}{nodelayer}
		\node [style=none] (1) at (-4, -4) {};
		\node [style=none] (2) at (4, 4) {};
		\node [style=none] (3) at (-3, -3) {};
		\node [style=none] (4) at (3, 3) {};
		\node [style=none] (5) at (0, 0) {};
		\node [style=none] (6) at (4, -4) {};
		\node [style=none] (7) at (-3, 3) {};
		\node [style=none] (8) at (1, 1) {};
		\node [style=none] (9) at (-2.5, 4.5) {};
		\node [style=none] (10) at (-2.25, 4.075) {};
		\node [style=none] (11) at (-2.25, 2.475) {};
		\node [style=none] (12) at (-2.825, 3.425) {$\Delta t$};
		\node [style=none] (13) at (-4, -3) {$X$};
		\node [style=none] (14) at (4, -3) {$X$};
		\node [style=none] (15) at (-3, -2.25) {};
		\node [style=none] (16) at (-2, -1.25) {};
		\node [style=none] (17) at (3, -2.25) {};
		\node [style=none] (18) at (2, -1.25) {};
		\node [style=none] (19) at (-2.75, -1.5) {$p$};
		\node [style=none] (20) at (2.95, -1.425) {$p'$};
	\end{pgfonlayer}
	\begin{pgfonlayer}{edgelayer}
		\draw [style=legarrow] (1.center) to (2.center);
		\draw [style=B] (3.center) to (4.center);
		\draw (6.center) to (5.center);
		\draw [style=R1] (5.center) to (7.center);
		\draw [style=legarrow] (8.center) to (9.center);
		\draw [style=arrow] (10.center) to (11.center);
		\draw [style=axis] (15.center) to (16.center);
		\draw [style=axis] (17.center) to (18.center);
	\end{pgfonlayer}
\end{tikzpicture}}
\caption{Schematic representation of the time delay.
}
\label{fig:t-delay}
\end{figure}

The time delay $\Delta t= -iS^*\partial_p S$ \cite{eisenbud1948formal,Wigner:1955zz} measures the extra time that it takes for a scattered particle to arrive to infinity, with respect to the time it would take if the particle were moving freely. In elastic $2\to 2$ scattering, we can treat one of the particles as the scatterer and focus on the time delay in the trajectory of the other particle (see figure \ref{fig:t-delay} for an illustration). In the case of massless particles, since they move on the edge of the lightcone, causality requires that the scattered particle only gets delayed, i.e.\ $\Delta t\geq 0$. This is a powerful condition which has been used in the past, for example, to constrain low-energy parameters in effective field theories \cite{Adams:2006sv,Camanho:2014apa}.

At the same time, it is well known that causality implies analyticity of the S-matrix. We thus might expect the positivity of the time delay to be already encoded in the analytic properties of $S$. A hint towards this comes from the fact that the bounds in \cite{Adams:2006sv,Camanho:2014apa} can be recovered from dispersion relations and positivity bounds \cite{Caron-Huot:2022ugt,Albert:2022oes}; one of the recent reincarnations of the S-matrix bootstrap leveraging analyticity and unitarity of scattering amplitudes.

In this appendix we provide a proof that $\Delta t\geq 0$ follows from standard assumptions about the S-matrix for the case of elastic scattering of massless particles in two dimensions. We comment on possible generalizations towards the end. We make the following (standard) assumptions about the S-matrix $S(s)$: (\textit{i}) it is analytic in the upper-half plane (UHP), (\textit{ii}) it is crossing symmetric and real-analytic, so that $S(-s^*) = S(s)^*$, (\textit{iii}) it saturates unitarity on the real axis, i.e.\ $|S(s)|=1$ ($s\in \mathbb R$), and (\textit{iv}) it is polynomially bounded as $|s|\to \infty$ in the UHP. Let us note immediately that (\textit{i})-(\textit{iv}) imply, via the Phragmén–Lindelöf principle, the stronger condition (\textit{iv}') $|S(s)|\leq 1$ in the UHP.

With these assumptions, when we map the UHP to the unit disk $\mathbb D$ with $s\to z\equiv \frac{i-s}{i+s}$, we get an analytic function on $\mathbb D$, bounded by $|S(z)|=1$ at $|z|=1$.
Such a function is known as an {inner function}. There exists a factorization theorem (see e.g.\ \cite[Ch.~II.6]{Garnett}) which states that every inner function factorizes into
\begin{equation}
    S(z) = B(z) e^{i2\delta_0(z)}\,,
\end{equation}
where $B(z)$ (known as a {Blaschke product}) captures all the zeros of $S(z)$ in the unit disk, and $e^{i2\delta_0(z)}$ (known as a {singular inner function}) is free of any such zeros. In terms of $s\in \text{UHP}$, the Blaschke product reads
\begin{equation}
    B(s) = \prod_j \frac{s_j - s}{s_j^* - s}\,, \quad (\text{Im}\,s_j>0)\,, \label{eq:Blaschke}
\end{equation}
where the collection of zeros can be infinite but must satisfy the {Blaschke condition} $\sum_j \text{Im}\,s_j/(1+|s_j|^2)<\infty$ so that (\ref{eq:Blaschke}) converges. Physically, these are the CDD factors in \eqref{eq:SCDD}, and condition (\textit{ii}) again requires them to be purely imaginary or come in pairs $s_{1j}^* = -s_{2j}$.

The phase shift factor $e^{i2\delta_0(s)}$ describes an elastic resonanceless scattering, and it is therefore analytic in the UHP. Crossing symmetry (\textit{ii}) implies $2\delta_0(-s^*) = -2\delta_0(s)^*$, while unitarity (\textit{iii}) sets $\text{Im}\, 2\delta_0(s) = 0$ on the real $s$-axis. We can derive a dispersive representation for this phase shift. We start by writing
\begin{equation}
    2\delta_0(s) = \frac{1}{2\pi i}\oint_{t\sim s\hspace{-5pt}} dt\left(\frac{1}{t-s} - \frac{1}{t+s}\right)2\delta_0(t)\,,
\end{equation}
and we blow out the contour in a sunrise shape around the UHP. For the integral on the real axis, since the kernel is symmetric under $t\leftrightarrow -t$, we can symmetrize the phase shift and exploit (\textit{ii}) and (\textit{iii}) to write
\begin{equation}
    \frac{1}{2}(2\delta_0(t+i\epsilon)+2\delta_0(-t+i\epsilon)) = i\, \text{Im}\, 2\delta_0(t)=0\,.
\end{equation}
So the real axis drops out, and the phase shift is completely determined from the arc at infinity.

For that arc, we make the change of variables $t \to -1/t$ to map the contour to a small clockwise arc around the origin.
By pushing the contour towards the real axis and repeating the symmetrization argument, we then obtain the desired dispersive representation:
\begin{equation}\label{eq:disp-delta}
    2\delta_0(s) = \frac{1}{\pi}\int_{-\epsilon}^{\epsilon} dt\, \frac{s}{1-(ts)^2}\,\text{Im}\, 2\delta_0(-1/t)\,.
\end{equation}
Note that $\text{Im}\, 2\delta_0(-1/t)\geq0$ as follows from $(iv)'$. To illustrate these points, consider the example $S(s)=e^{is}$, for which we have $\text{Im}\, 2\delta_0(-1/s) = \pi \delta(s-0)$; a positive distribution. In contrast, for the case $S(s)=e^{is^3}$, which is not bounded in the UHP, we have $\text{Im}\, 2\delta_0(-1/s) = \frac{\pi}{2} \partial_s^2\delta(s-0)$. The representation~\eqref{eq:disp-delta} still works here, but the measure is not positive.

We are finally ready to show the positivity of the time delay. Taking the $p$-derivative (where $s=4pp'\geq 0$),
\begin{equation}
    \Delta t = 4p' \frac{\partial_s S(s)}{iS(s)} = 4p' \left(\frac{\partial_s B(s)}{iB(s)} + \partial_s 2\delta_0(s)\right)\,.
\end{equation}
The Blaschke product contributes positively simply because zeros are in the UHP;
\begin{equation}
    \frac{\partial_s B(s)}{iB(s)} = \sum_j \frac{2\, \text{Im}\,s_j}{|s_j-s|^2}\geq 0\,.
\end{equation}
The resonanceless phase shift also contributes positively due to the positivity of the measure;
\begin{equation}
    \partial_s 2\delta_0(s) = \frac{1}{\pi}\int_{-\epsilon}^{\epsilon} dt\, \frac{1+(ts)^2}{(1-(ts)^2)^2}\,\text{Im}\, 2\delta_0(-1/t)\geq 0\,.
\end{equation}
We conclude that $\Delta t\geq 0$ for all energies follows from the assumptions (\textit{i}-\textit{iv}) listed above.

The intuitive argument for $\Delta t\geq 0$ sketched at the beginning of this appendix (and depicted in figure \ref{fig:t-delay}) is difficult to make rigorous. It requires working carefully with wavepackets so that we can discuss the trajectory of the particles. This complication has led some authors to question the positivity of $\Delta t$ (see e.g.\ \cite{Chen:2023rar}). Our proof bypasses entirely the need for wavepackets, and establishes $\Delta t\geq 0$ from basic assumptions of the S-matrix, settling --we hope-- this debate. Of course, our proof applies only to $2d$ massless elastic scattering. Dropping each of these adjectives is an outstanding research direction on its own:
\begin{itemize}
    
\item For inelastic scattering, a generalization of the notion of time delay exists, known as the Wigner-Smith operator \cite{Smith:1960zza,Martin:1976iw},
    \begin{equation}
        \widehat{\Delta t} \equiv -\frac{i}{2}\left(\hat S^\dagger \partial_{E} \hat S + \partial_{E} \hat S^\dagger \hat S\right)\,.
    \end{equation}
It would be remarkable to show that this operator is positive-semidefinite. The jets of \cite{Guerrieri:2024ckc} could be useful in this regard. 
    
\item Going to higher dimensions inevitably requires understanding the inelastic setup as there particle production is mandatory \cite{Aks:1965qga}. Of course, in semiclassical regimes where the amplitude exponentiates to large phases, time delay constrains are well understood in any dimension \cite{Camanho:2014apa,Arkani-Hamed:2020blm}. The discussion above suggests a stronger result may be obtained.

\item  Massive particles do not travel at the edge of the lightcone. They could therefore suffer some time advance without violating causality. It would be interesting to find a lower bound on the allowed time advance. The discussion around (\ref{eq:DMB}) could be relevant for that.
\end{itemize}

\vspace{-0.6cm}
\section{Ruling out the integrable zig-zag model}
\label{app:zigzag}

In $D=3$, the simplest worldsheet S-matrix of a single transverse Goldstone compatible with non-linearly realized Poincaré is the shockwave S-matrix $    S(s)= e^{i c s}$, where $c>0$ so that the S-matrix is bounded in the upper half-plane, see appendix \ref{app:criticalST}. It describes the scattering of Goldstones on a long Nambu-Goto string. The asymptotic linear behavior of the phase shift has been proposed as the dominant contribution on the Goldstone scattering in gauge theories as well \cite{Dubovsky_2018}, based on a semiclassical picture of a zig-zag forming during a high energy worldsheet scattering.

The purpose of this section is to show that an integrable worldsheet S-matrix whose phase shift grows as $\log S \sim i cs$ at large $s$ cannot, for any choice of R-matrix, reproduce the Coulombic logarithmic singularity of the open string ground state energy at small $\tau$, equation (\ref{eq:perturbativeasymp}). We assume that the Goldstone is the only stable particle and that the potential is smooth in $\tau$. 

The argument is simple. Using boost invariance of the S-matrix, we obtain, from the TBA equation (\ref{eq:eps-eq}), that the pseudoenergy $\eps(p)$ is linear in $p$ at large $p$:
\begin{equation}
    \eps(p) \sim p \left(2 \tau + 4c \int_0^\infty \frac{dp'}{2\pi} \log\left(1-|K(p')|^2 e^{-\eps(p')}\right)\right).
\end{equation}
Note that the onset of this behavior is controlled by the asymptotics of the S-matrix, which does not depend on $\tau$. Moreover, we recognize that the integral has reduced to the ground state energy (\ref{eq:E-eq}), so that 
\begin{equation}
    \eps(p) \sim p \left(2  \tau + 4 c \left(V(\tau) - \ell_s^{-2}\tau\right)\right).
\end{equation}

We now assume that the potential diverges logarithmically as $V(\tau)\sim \frac{\lambda}{4\pi} \log \tau$ at small $\tau$. We see that, at small but finite $\tau$ (and for sufficiently large $p$), the pseudoenergy depends linearly on $p$ with a large negative coefficient that can be made larger by further reducing~$\tau$. We thus find that, for any non-trivial $K(p)$, the asymptotic behavior of $|K(p)|^2 e^{-\eps(p)}$ can be made arbitrarily large.
This forces a singularity of the integral in (\ref{eq:E-eq}), forcing in turn a singularity of the potential at finite $\tau$; a contradiction. Therefore, the assumption of a logarithmic diverging potential which is smooth at $\tau >0$ is incompatible with the linearly growing phase shift.

The intuition behind this result is that the linear phase shift at large $p$ amounts to such strong (attractive) interactions that, in integrability, it lowers the energy levels to a point where the partition function diverges. Note, however, that we make no claims of excluding such a phase shift
in the actual Yang-Mills string. There, high energy goldstone scattering is expected to be accompanied by a shower of soft Goldstones~\cite{Dubovsky_2018}, thus making the process highly non-integrable. Finally, let us conclude by pointing out that the critical string worldsheet precisely realizes a linearly growing phase-shift, and its corresponding open string energy suffers from a finite $\tau$ singularity (due to the open string tachyon), see appendix~\ref{app:criticalST}.

\end{document}